\documentclass[fleqn,usenatbib]{mnras}
\usepackage{float}
\usepackage{varioref}

\usepackage[T1]{fontenc}
\usepackage{ae,aecompl}

\usepackage{latexsym}
\usepackage{graphicx,epsf,epsfig} 
\usepackage{amssymb,amsmath}
\usepackage{bm}
\usepackage[usenames]{color}
\usepackage{pstricks}
\usepackage{rotating}
\usepackage{feynmf}
\usepackage[T1]{fontenc}

\usepackage{graphicx}
\usepackage{dcolumn}
\usepackage{bm}
\usepackage{color} 

\title[ExSHalos: A New Parameter Free Method for Fast Generation of Halo Catalogues]{Excursion Set Halos - ExSHalos: \\ A New Parameter Free Method for Fast Generation of Halo Catalogues}

\author[R. Voivodic, M. Lima and L. R. Abramo.]{
Rodrigo Voivodic,$^{1}$\thanks{e-mail: rodrigo.voivodic@usp.br}
Marcos Lima$^{1}$
 and L. Raul Abramo$^{1}$
\\
$^{1}$Departamento de F\'{\i}sica Matem\'atica, Instituto de F\'{\i}sica, Universidade de S\~ao Paulo, CP 66318, CEP 05314-970,  S\~ao Paulo, SP, Brazil \\
}

\date{Accepted XXX. Received YYY; in original form ZZZ}

\pubyear{2019}

\begin{document}
\label{firstpage}
\pagerange{\pageref{firstpage}--\pageref{lastpage}}
\maketitle

\begin{abstract}
We develop a new, simple, fast and parameter-free method to construct dark matter halo catalogues. This method requires as inputs only the linear matter power spectrum and the threshold density for halo formation in linear theory. It directly uses excursion set ideas and Lagrangian perturbation theory to produce halo catalogues with the correct abundance, large scale power spectrum, bispectrum and velocity field. These halo catalogues can be used for the fast construction of mock galaxy catalogues, allowing for the evaluation of covariance matrices for multiple observables. Because of its robustness and predictive nature, this method can be easily adapted to produce catalogues with e.g. primordial non-Gaussianities, modified theories of gravity and non-standard dark energy models, enabling detailed studies of these models in the context of next-generation surveys. We implement this method in a \texttt{C} code, and present numerical comparisons with theoretical predictions as well as full N-body cosmological simulations.
\end{abstract}


\maketitle

\section{Introduction}

Within the Halo Model for structure formation \citep{Cooray}, dark matter halos are the skeletons upon which luminous matter such as galaxies flesh out the universe large-scale structure \citep{Zheng}. 
Starting off from N-body simulations of dark matter particles, we may run halo finders and extract catalogues of dark matter halos. A Halo Occupation Distribution (HOD) model provides a prescription for how galaxies populate these halos, allowing for the construction of simulated galaxy catalogues \citep{Berlind, Berlind2}. These catalogues are used in multiple applications in Astrophysics and Cosmology, and a large number of them are frequently required for a precise estimation of statistical quantities \citep{Manera, Manera2, Avila2}.

For instance, galaxy catalogues may be constructed expressly for the purpose of studying the properties of a particular galaxy survey \citep{Manera}. In this case, a number of codes can be run and calibrated on the simulated catalogues, including estimation of N-point correlations functions, power spectra, cluster finders, codes for photometric redshifts, etc. Likewise, thousands of these simulated catalogues can be used to accurately estimate auto- and cross-covariance matrices relating multiple observables \citep{Blot2, Lippich, Colavincenzo}. 

However, the very first step in this chain, i.e. running accurate and high-resolution N-Body simulations, is usually a slow process that requires intensive use of parallel computational resources \citep{Teyssier, Springel}. The next step of finding halos according to some prescription can also be relatively slow and computationally intense \citep{Knebe}. As a result, in many cases the production of thousands of independent simulated galaxy catalogues may just be unfeasible, due to these computational bottlenecks. Given this limitation, a number of recent works have focused on the production of fast and accurate dark matter halo catalogues, without the need of running full N-body simulations and halo finders.

As discussed in the three papers of the comparison project ``Comparing approximate methods for mock catalogues and covariance matrices'' \citep{Blot2, Lippich, Colavincenzo}, there are three classes of approximate methods to generate halo catalogues: i) methods with free parameters that need to be fitted using simulations, ii) simple recipes that do not use free parameters but reproduce only some observables, and iii) predictive methods that do not require fits to simulations but, in some cases, have internal parameters that can be specified in order to improve performance.

The first class, which requires fits to N-body simulations, displays high accuracy and great performance, enabling the generation of a large number of halo catalogues. On the other hand, these methods also require large computational time to ensure that the cosmological model under consideration is faithfully realized by simulations. For each model, typically a small set of N-body simulations is produced, along with their halos, and in each case free parameters need to be refitted. As a result, the study of dark energy, modified gravity and primordial non-Gaussianities is often impractical with these methods, since the parameter space is too large to be properly explored. Examples of methods in this class include PTHalos \citep{Scoccimarro}, HALOGEN \citep{Avila}, EZmock \citep{Chuang2} and PATCHY \citep{Kitaura2}.
In general these methods require free parameters for defining (or finding) halos in the density grid generated using Lagrangian Perturbation Theory (LPT), because the parametrizations usually employed in N-body simulations (e.g. spherical overdensity $\Delta = 200$) do not work \citep{Scoccimarro}.

The main example of the second class is the lognormal method, which consists in generating a density map taking the exponential of the Gaussian density field, and then using this map to place galaxies in a box following a Poisson distribution \citep{Coles,Abramo2015,Xavier2016}.
This approach is simple, fast and does not require free parameters. However, it does not incorporate properties of halos and their matter distribution, i.e. there is no information about nonlinear scales. As a result, the method accurately recovers the linear power spectrum on large scales, but fails to properly reproduce small scales and higher order correlations.


In the third class we have predictive methods, which do not require N-body simulations, but still provide the correct statistics with some precision. These methods are typically less accurate compared with the first class of methods. 
The methods in this class have some free parameters that need to be set in order to improve results and efficiency, but that do not need to be refitted every time the cosmology is changed. Some examples of these methods are COLA \citep{Tassev} and fastPM \citep{Feng2}, which run a low resolution simulation along with LPT. The PINOCCHIO algorithm \citep{Monaco} finds halos in the initial Gaussian density grid solving the ellipsoidal collapse equations and, more recently, the mass-Peak Patch algorithm \citep{Stein} works similarly. An extended comparison between methods, along with a discussion and details about the pros and cons of each one, is presented in \cite{Chuang3}.

In this work we propose a new, fast and parameter-free method to construct dark matter halo catalogues. We refer to it as the Excursion Set Halos (ExSHalos) method. Our method lies between the second and the third class of methods described above: on one hand, like the second class, our method has no free internal parameters, but on the other hand, the main ideas behind the method, as well as the results, are close in spirit to those of the third class of models. Our method allows for the possibility of changing the barriers for halo formation by tuning some free parameters, which makes it more similar to the third class method. In particular, ExSHalos is similar to PINOCCHIO \citep{Monaco} and the mass-Peak Patch algorithm \citep{Stein} in the way that it finds halos in Lagrangian space. The main difference of our method is that we do not solve the ellipsoidal collapse equation to determine the time of collapse of some smoothed region. Instead, we measure directly the "trajectory"\footnote{Here trajectory has the same meaning as in theoretical computations, in the sense of a relation between the mean density inside a sphere and its radius.} around peaks, and determine the position of the first crossing through the linear halo barrier.

Our method relies strongly on ideas from the Excursion Set approach for halo formation \citep{Bond}, and requires only two inputs to produce halo catalogues: a linear matter power spectrum and a prescription for the overdensity threshold for halo formation in linear theory. The halo catalogues produced by this method have the correct halo properties on large scales, with (1) the correct halo abundance or mass function, (2) the correct halo bias, and (3) a fair reproduction of the shape of the power spectrum and bispectrum. The resulting halo catalogs are accurate enough for purposes of constructing simulated galaxy catalogues based on HODs or semi-analytical prescriptions. This allows for the creation of thousands of independently generated catalogues in a few minutes using only a laptop computer, making possible the computation of realistic and accurate covariance matrices. Since this method is based on simple and general ideas of halo formation (namely, the Excursion Set approach), it can be easily applied on cosmological models different from those tested in this work. Because of these general aspects, the method can also be modified to produce halo and galaxy catalogues with modified theories of gravity, dark energy and primordial non-Gaussianities, something which nontrivial to achieve using other methods. 


This paper is organized as follows. 
In \S~\ref{sec:algorithm} we describe in detail the algorithm proposed to generate dark matter halo catalogue realizations starting from a dark matter linear power spectrum. We also review some elements of the Excursion Set approach, as well as Lagrangian Perturbation Theory (LPT), which are the main theoretical elements of the method. 
In \S~\ref{sec:Comparions_Outputs} we analyze the results of our method for a given cosmology, testing different barrier choices and LPT order, performing a number of robustness tests. 
In \S~\ref{sec:Comparisons_Simulations} we contrast the results of our method to those of full N-Body simulations, comparing a number of halo statistical properties. 
Finally, in \S~\ref{sec:Conclusions} we present our main conclusions. 
The Appendices at the end present a few variations of the fiducial prescription, additional robustness tests of the algorithm, and a one-to-one comparison with a low resolution simulation.

\section{Algorithm Description}
\label{sec:algorithm}

	In this section we describe each step in the algorithm used to generate halo catalogues, along with their theoretical underpinnings. The consistency checks, comparisons with theoretical predictions, checks of the performance of our current implementation and a full comparison with N-body simulations are described in the next sections.
    
	The algorithm is based on the main ideas of the Excursion Set theory \citep{Bond,Maggiore1}, along with  Lagrangian perturbation theory \citep{Vlah, Matsubara2}, and consists of three main steps:
    \begin{itemize}
    \item The generation of a linear Gaussian density distribution, in a grid, following a given linear matter power spectrum;
    \item The halo finding stage, where halos are located on the grid according to a density threshold for halo formation in linear theory, i.e. the halo barrier;
    \item The halo displacement stage, where halos are moved to Eulerian space using Lagrangian perturbation theory, thus correcting their correlation functions.
    \end{itemize}
    
    The next subsections describe each of these steps in detail.
    
\subsection{Generating the Gaussian Field}

	The first part of the algorithm is to generate the linear density field in terms of a Gaussian random field on a grid, according to a given input {\em linear} matter power spectrum. 
    
    We start by constructing a cubic grid with some physical size $L$ and with a given number of divisions per dimension $N_{d}$\footnote{Note that the choice of these numbers and the value of the background matter density $\Omega _{m}$ fix the minimum mass of each halo in the final catalogue.}. For a given cosmology, we compute the linear matter power spectrum $P_{L} (k)$ using the Einstein-Boltzmann solver code \texttt{CAMB} \citep{CAMB}. We generate a realization of this spectrum by sampling random complex numbers from a Gaussian distribution (using the Box Muller transform) in the Fourier-space version of the grid. We impose the reality condition of the density field in configuration space by setting $\delta ^{*} (\textbf{k}) = \delta (-\textbf{k})$. The distribution of this random density field in Fourier space will be:
    
\begin{equation}
\mathcal{P}[\delta (\textbf{k})] = \frac{1}{\sqrt{2\pi P_{c}(k)}} \exp \left[ - \frac{\left| \delta (\textbf{k}) \right| ^{2}}{2 P_{c}(k)} \right] \,, 
\label{eq:Gaussian}
\end{equation}
where $P _{c}(k)$ is the convolved power spectrum given by
\begin{equation}
    P_{c}(k) = 4 \pi \int _{0} ^{L/2} dr \frac{\sin (kr)}{kr} r^{2} \xi _{L}(r) \, . 
\label{eq:renPower}    
\end{equation}
Here $\xi _{L} (r)$ is the linear matter correlation function, which is evaluated in terms of the linear matter power spectrum:
\begin{equation}
    \xi _{L}(r) = \frac{1}{2 \pi ^{2}} \int _{0} ^{\infty} dk \frac{\sin (kr)}{kr} k^{2} P_{L} (k) \, .
\label{eq:Power}
\end{equation}

    We use the convolved power spectrum in order to retain the correct statistics in real space, which is where we will look for halos and construct our catalogue. A complete discussion about the use of the convolved power spectrum to generate initial conditions for N-body simulations can be found in \cite{Sirko}. To compute the integrals in Eqs.~\eqref{eq:renPower} and \eqref{eq:Power} we use the \texttt{FFTLog} algorithm \citep{Hamilton}.
    
   Next we take the inverse Fourier transform of the grid to obtain a real-space density grid with the correct two-point correlation function and vanishing higher-order correlation functions. This procedure gives to us a realization of the linear theory prediction for the matter distribution in the universe for a given cosmology and redshift. This is similar to the generation of initial conditions for N-body simulations.
    
    At the end of this process we have $N_{d}^{3}$ real values of the Gaussian density field $\delta _{ijk}$ (with $i, j, k = 0, 1, ..., N_{d}-1$) evaluated at points $(x,y,z)=(iL/N_{d} + L/(2N_{d}), jL/N_{d} + L/(2N_{d}), kL/N_{d} + L/(2N_{d}))$, whose two-point function is the one we selected on the basis of our cosmological model. We note that, in principle, this density field has values between $-\infty$ and $+\infty$, which makes it impossible to use directly to generate mock catalogues. This is the motivation for theoretical transformations, such as those used in lognormal maps \citep{Coles}. 

\subsection{Finding Halos}

	Once the Gaussian density grid has been constructed, we apply the ideas from the excursion set theory in order to find/define dark matter halos. In this context, one key hypothesis is that the information about halos is fully encoded in the initial conditions. Therefore, it should be possible to find all halos and their masses by looking only at the initial conditions -- i.e. that information is already encapsulated in the initial Gaussian density field.
    
    First, we look at all cells in the grid and find those that are density peaks, in such a way that the densities of all their neighbors are smaller. These density peak cells are natural halo center candidates. These peaks will be the centers of future halos, unless some peak happens to lie inside another larger halo, or in case that peak does not surpass the density threshold.
    
    With the positions of the halo centers we need to associate a mass to each of them. To do this we grow a sphere around each peak, joining the neighbouring cells until the mean density inside the sphere becomes smaller than some threshold, which in general will be a function of the halo size. At present we have implemented two different thresholds (barriers) for halo formation:    
    \begin{eqnarray}
    B_{SB}(S) &=& \delta _{c} \,, \\
    B_{EB}(S) &=& \sqrt{a}\delta _{c} \left[ 1.0 + \beta (a\nu ^{2})^{-\alpha} \right]  \,,  
    \label{eq:Barriers}
    \end{eqnarray}
    where $B_{SB}$ is just the constant barrier used to compute the halo mass function in \cite{Press} and \cite{Bond}, and $B_{EB}$ is an approximation for the ellipsoidal collapse barrier \citep{ST, Sheth2, deSimone2}. Here $\delta _{c}$ is the linear density contrast for spherical collapse and $a$, $\beta$ and $\alpha$ are free parameters\footnote{In this work we set $a=0.72$, $\alpha = 0.98$ and $\beta = 0.36$ in such a way to reproduce the large scale power spectrum measured from our simulation (see \S~\ref{sec:Comparisons_Simulations}). 
    }. In Fig.~\ref{fig:Halo_Barriers} we show the mass dependence of these two barriers. The peak-height $\nu = \delta _{c}/\sigma (R)$ is given in terms of $S$, which is the variance of the linear density field and is related with the halo radius $R$ through 
	\begin{equation}
  	S(R) = \sigma^2(R) = \int _{-\infty} ^{\infty} \frac{dk}{2\pi ^{2}} k^{2} P(k) |W(kR)|^2 \,,  
    \label{eq:Variance}
    \end{equation}
    where $W(kR)$ is some window function used to smooth the density field at a scale $R$, typically considered as a top-hat function.

\begin{figure}
	\begin{center}
	\includegraphics[width=\linewidth]{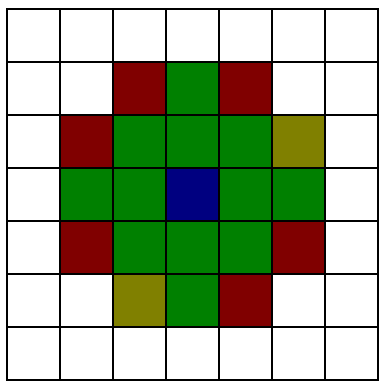}
	\caption{A 2D representation of the halo finder in the Gaussian grid. The blue cell at the center represent the current peak around which we grow a "sphere", the green cells represents a "sphere" where the density is still larger than the threshold barrier and the red + yellow cells represents the next "sphere" where the density is smaller than the barrier. The yellow cells are put in the halo to improve the mass resolution due to discretization.}
	\label{fig:Halo_Finder}
	\end{center}
	\end{figure}

    
    At the end of this stage we have approximate spheres consisting of the union of grid cells. In Fig.~\ref{fig:Halo_Finder} we show a 2D example of one such a sphere, grown around a peak. The blue cell represents the peak cell, while the green cells (together with the blue) represent the sphere where the mean density is larger than the barrier threshold with radius of $2L_{c}$, where $L_{c} = L/N_{d}$ is the length of one cell. The red cells represent the next sphere, with radius $\sqrt{5}L_{c}$, which has a mean density smaller than the barrier, and the yellow cells represent additional cells which are included in the halo in order to improve its mass resolution. In this example we found a halo with center at the position of the blue cell, and with a mass given by the total volume of the blue + green + yellow cells times the background matter density (in this case, $M_{h} = 15 \times L_{c}^{2} \times \bar{\rho} _{m}$).
    
	We include the yellow (adjacent) cells in order to improve the mass discretization of halos. If we had considered only spheres, our halos would end up with only 1, 7, 19, $\ldots$ cells, since in 3D we have 1 cell in a sphere of radius $L_{c}/2$, 7 cells in a sphere of radius $L_{c}$, 19 cells in a sphere of radius $\sqrt{2} L_{c}$, and so on. In order to select these yellow cells we consider all cells in the next shell: first we sort the cells by density, then we start incorporating them, one by one, in order of density, until the mean halo density becomes smaller than the barrier, or until we reach a cell that belong to another halo.
	The final halos will not be perfectly spherical, but they will have masses which are given by any integer multiple of $L_{c}^{3} \, \bar{\rho} _{m}$.
    
    Note that, in the case of a moving barrier, it is possible that it can be crossed twice. In order to take into account only the largest halo, we grow the sphere until its mean density is smaller than the minimum value of the barrier (we expect that the barrier decreases with halo mass), then we shrink the sphere until its mean density is higher than the barrier value. This procedure ensures, as in the excursion set theory, that we take the first value of $S$ that crosses the barrier, starting from large scales and going to small ones. Since typically the mean density around a peak decreases monotonically with volume, we expected that there is not a second crossing in the constant barrier case.
    
    \begin{figure*}
	\begin{center}
	\includegraphics[width=\linewidth]{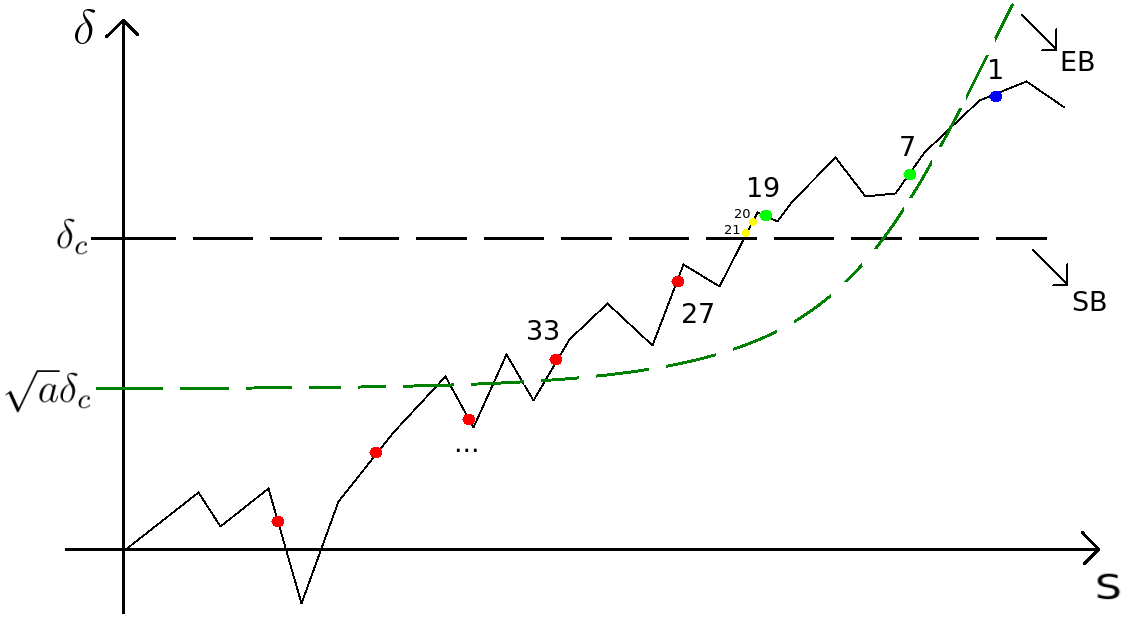}
	\caption{Another representation of our halo finder in comparison with the standard excursion set approach. We show the "trajectory" made by one halo (black solid line), along with some dots displaying the number of cells in each sphere grown, which represent our only information about the trajectory. The colors of dots are the same as in Fig.~\ref{fig:Halo_Finder} with respect to the static barrier (black dashed line). We also plot the ellipsoidal barrier in green dashed line.}
	\label{fig:Excursion}
	\end{center}
	\end{figure*} 
	
	In Fig.~\ref{fig:Excursion} we represent how the ``trajectory'' of a halo will be discretized by our method. The black solid line represents the trajectory of some halo, and the dots show where we have access to that information. The colors are the same of  Fig.~\ref{fig:Halo_Finder}, with the blue dot indicating the peak cell, the green dots representing the shells above the barrier, the red dots showing the shells below the barrier, and the yellow dots indicating the cells included in the halo so as to improve its mass discretization. The colors were chosen taken as reference the static barrier (black dashed line). This figure also shows how the yellow cells improve the mass resolution of our halos, since without them all halos between 19 and 27 cells would be grouped together.
	
    Now looking at the ellipsoidal barrier in Fig.~\ref{fig:Excursion} (green dashed line), we see the importance of taking care of the double crossing. In this particular example, if we considered the barrier and not its minimum as we grow the sphere, we would not have a halo because the density in the center (blue dot) is below the barrier. In our case, as we grow the sphere up to the minimum of the barrier ($\sqrt{a} \, \delta _{c}$), we find this halo and, within our lattice resolution, we recover the correct first crossing scale.

	
	With the ellipsoidal barrier our method becomes more akin to the third class of methods, since we have three parameters that can be tuned to improve the results. A comparison between the properties of the final halo catalogue with different choices of barrier is presented in Appendix~\ref{app:Variation}.

\subsection{Displacing Halo Centers}

	The position of each halo will be associated with the position of one grid cell, the central (blue) one. However, we know that the position of the particle associated with that cell will be moved due to gravitational interactions. Therefore, we should be able to move this particle, as well as its parent halo, to its correct place. To do this we use Lagrangian Perturbation Theory (LPT), which displaces the initial (Lagrangian) particle positions to their final (Eulerian) positions: 
\begin{equation}
\textbf{x}(\textbf{q}) = \textbf{q} + \textbf{s}(\textbf{q})\, ,
\label{eq:LPT}
\end{equation}
where $\textbf{q}$ is the initial (Lagrangian) position of the particles (which is the center of some grid cell), $\textbf{s}$ is the displacement field, and $\textbf{x}$ is the final (Eulerian) coordinate of the particle.

    We assume that this prescription is sufficient to assign halos to their correct positions, since it is not expected that the halos have very large displacement fields. This statement is justified by, e.g., the good agreement between the excursion set theory linear bias and the linear bias measured from N-body simulations \citep{Lazeyras, Tinker2}.

	In this process we also assume that halos retain their original mass \cite[the main idea of][]{Press}, and that they are located at the positions of the central particles, associated with the peak cells. We expect this second assumption to be correct up to the limit of validity of LPT.
	  
    Up to second order in LPT, the Eulerian position of the particles, as well as their velocities, are given by \citep{Scoccimarro}
\begin{eqnarray}
\label{eq:2LPT_pos}
\textbf{x}(\textbf{q}) &=& \textbf{q} - D_{1} \nabla _{q} \phi ^{(1)} + D_{2} \nabla _{q} \phi ^{(2)}\,, \\
\textbf{v}(\textbf{q}) &=& - D_{1} f_{1} H \nabla _{q} \phi ^{(1)} + D_{2} f_{2} H \nabla _{q} \phi ^{(2)}\,,
\label{eq:2LPT_vel}
\end{eqnarray}
where $D_{1}$ is the linear growth function, $D_{2}$ is the second-order growth function, which is well approximated by $D_{2} \approx - 7/3 D_{1}^{2}$ \citep{Bouchet}, $f_{i} = d \ln D_{i} / d \ln a$ are the growth rates, well approximated by $f_{1} \approx \Omega _{m} ^{5/9}$ and $f_{2} \approx 2 \Omega _{m} ^{6/11}$. In first order (1LPT) we keep only terms proportional to $D_1$, and to second order (2LPT), we keep all terms proportional to $D_1$ and $D_2$. Here $\phi ^{(1)}$ and $\phi ^{(2)}$ are the first- and second-order potentials given by
\begin{eqnarray}
\label{eq:Potential1}
\nabla _{q} ^{2} \phi ^{(1)} &=& \delta  \,, \\
\nabla _{q} ^{2} \phi ^{(2)} &=& \sum _{i>j} \left[ \phi ^{(1)} _{,ii} \phi ^{(1)}_{,jj} - (\phi ^{(1)}_{,ij})^{2}  \right] \, ,
\label{eq:Potential2}
\end{eqnarray}
where $\delta (\textbf{q})$ is the linear density field generated in step 1, where the halos were found. 
In Eqs. (\ref{eq:2LPT_pos})-(\ref{eq:Potential2}), the dependencies on $\textbf{q}$ and time are implicit in all quantities, and $_{,i} = \partial / \partial q_{i} $.

\begin{figure*}
	\begin{center}
	\includegraphics[width=\textwidth]{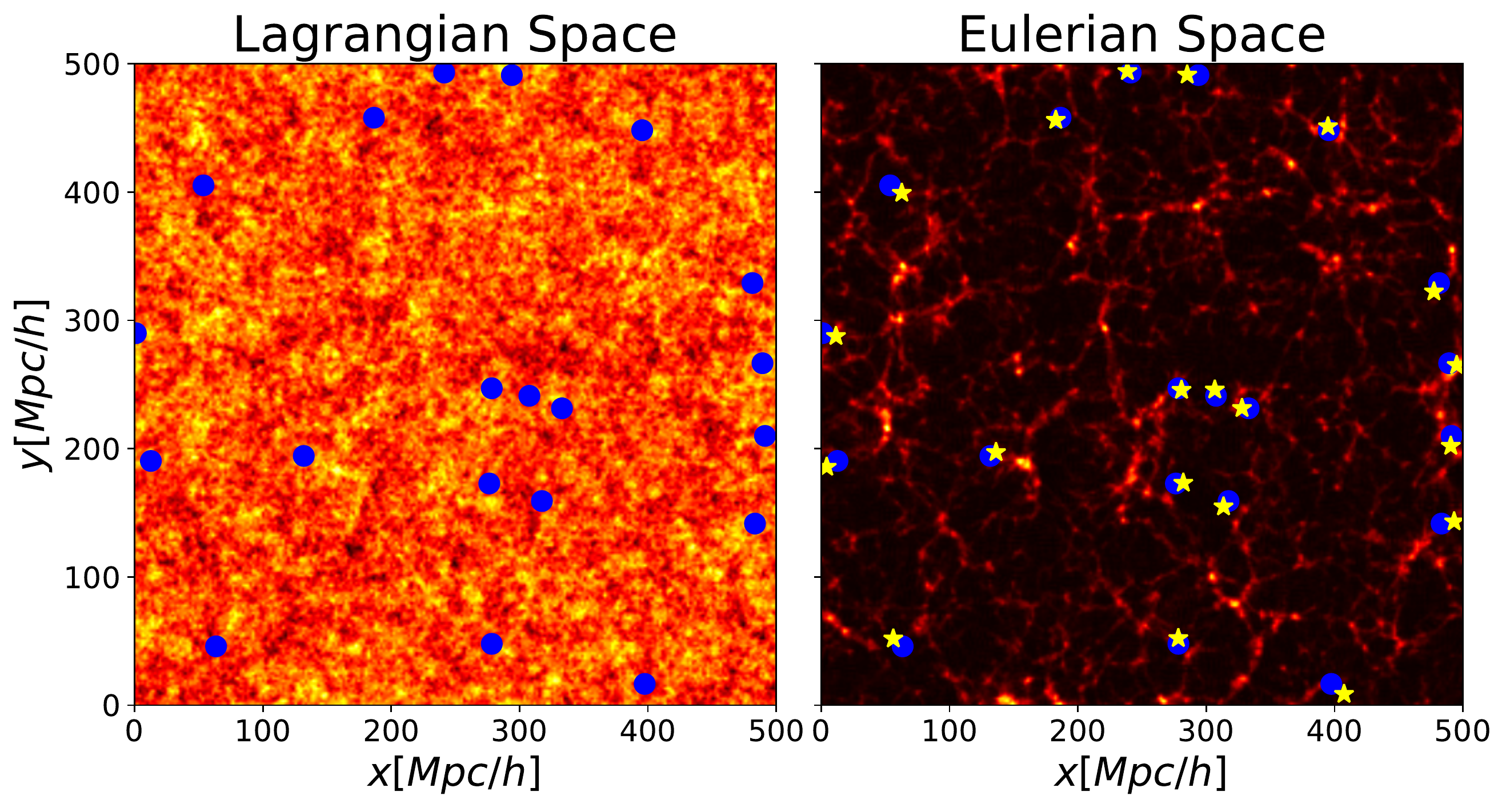}
	\caption{Slice of the density field for the Gaussian initial conditions ({\it left}), and the corresponding displaced 2LPT density map ({\it right}). The blue dots represent the 20 most massive halos found in this density slice, and the yellow stars are their positions after the 2LPT displacement.}
	\label{fig:Density_Map_LPT}
	\end{center}
	\end{figure*}

    In Fig.~\ref{fig:Density_Map_LPT} we show a slice of the matter Gaussian density map (left), and the same slice after the 2LPT displacement of the particles (right). We also show the $20$ most massive halos in this slice before (blue dots) and after (yellow stars) the displacement. In the left panel we can see that the halos are near the densest regions (shown in yellow), as expected. In the right panel the density map is less homogeneous, displaying some structures, like clumps and filaments. This occurs because 2LPT reproduces the two- and three-point functions, at linear order, and provides a good approximation to the higher order correlation functions as well \citep{Scoccimarro2, Bouchet}. Moreover, halos lie in the vicinity of dense regions, and their displacements are relatively small compared with the sizes of filaments.
    
    In Appendix~\ref{app:Variation} we show how the LPT order affects the halo density map, the two- and three- point correlation functions, as well as the two-point density-velocity and velocity-velocity correlation functions. A one-to-one comparison with a low resolution simulation using the same initial conditions is presented in Appendix~\ref{app:One}.

\section{ExSHalos Results}
\label{sec:Comparions_Outputs}

	In this section we analyze the main results 
	of our method. We compare the abundance and the linear bias of the halo catalogues generated using different barriers and different LPT orders. A comparison between the power spectrum, bispectrum and velocity spectrum for the different options is presented in Appendix~\ref{app:Variation}, along with a discussion on the effects of each option.
	
    We also show how the computational resources (i.e. time and memory) needed by our method depend on the box size and on the resolution (number of grid cells per dimension).

\subsection{Mass Function and Linear Bias}

	In order to compare the halo catalogues generated using the two different barriers Eq.~\eqref{eq:Barriers} and the three different LPT orders Eq.~\eqref{eq:LPT}, we generated eight halo catalogues for three different box sizes ($512, 1024$ and $2048$ Mpc$/h$) using $512$ cells per dimension in the grid. This allows us to proble a large range of halo masses. We fix the cosmology to that of the N-body simulations described in \S~\ref{sec:Comparisons_Simulations}.
	
	The catalogues used in the comparisons are listed in Table~\ref{tab:Catalogues_test}, as well as their main characteristics. In particular, note that the SB case, used in the next section, and  the 2LPT case correspond to the same catalogue.
    
\begin{figure*}
	\begin{center}
	\includegraphics[width=\linewidth]{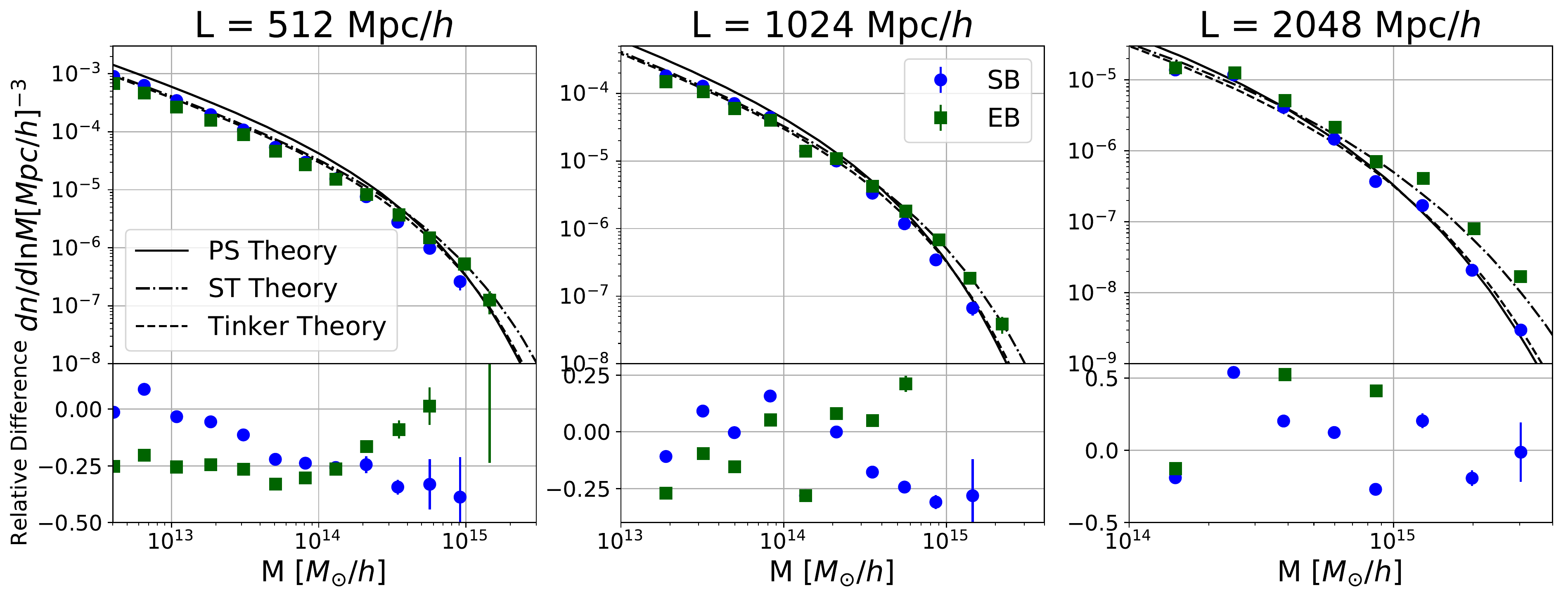}
	\caption{Mean linear halo mass function for eight realizations of each of the three box sizes generated using the SB barrier ({\it blue dots}) and the EB barrier ({\it green squares}) with $\pm 1\sigma$ errors. We also show theory predictions of \citep{Press} (PS, solid line), \citep{ST} (ST, dotted-dashed line) and \citep{Tinker} (Tinker, dashed line). The bottom panels show the relative differences of each catalogue with respect to Tinker's fitting formula. The panels show results for boxes with sizes $L=512$, $1024$ and $2048$ Mpc$/h$ from left to right.}
	\label{fig:Abundance_test}
	\end{center}
	\end{figure*}

    In Fig.~\ref{fig:Abundance_test} we show the mass function for the halos generated using the two different barriers and with some theory predictions. Each panel shows the results for each box size used. The points denote the mean over eight realizations and the error bars indicate $\pm 1 \sigma$ errors.
    
    We can see that the SB catalogue has a mass function within $25\%$ agreement with respect to the prediction of \cite{Tinker} (dashed line). Our EB catalogue has a better agreement with the prediction of \cite{Sheth2} (dotted-dashed line). 
    
    The agreement between our EB catalogue and the prediction of \cite{Sheth2} is expected since in this case the same barrier model was used to generate simulated halos and to compute this theoretical prediction. The result that the SB catalogue has a better agreement with the fit of \cite{Tinker} compared to the prediction of \cite{Press} comes from the fact that we do not need to truncate the expansion of the mass function in the non-Markovianities, as done in  Press-Schechter theory. Clearly, the mass function computed with a static barrier when all non-Markovian terms are taken into account produces a better result than the truncated expansion. For a discussion of non-Markovian corrections and the path integral formalism of the excursion set theory, see \cite{Maggiore1,Maggiore2,Maggiore3}. A comparison of different mass functions computed via the path integral formalism is presented in \cite{Corasaniti,Corasanitti2}.
    
    Our halo catalogues also recover the relative features between the two barriers, with fewer EB halos than SB halos at low masses and more halos at large masses. The transition between the two regimes happens at the expected mass scale, $\approx 10^{14} M_{\odot}/h$ as shown in Fig.~\ref{fig:Halo_Barriers}.
    
    Note that the small oscillations around the apparently smooth behaviour are not a statistical scatter and happen due to the nature of our method, which constructs halos as the union of cubic boxes. In particular, we point out that these oscillations are smaller in catalogues with large resolution (i.e. smaller cells). 
    
\begin{figure*}
	\begin{center}
	\includegraphics[width=\linewidth]{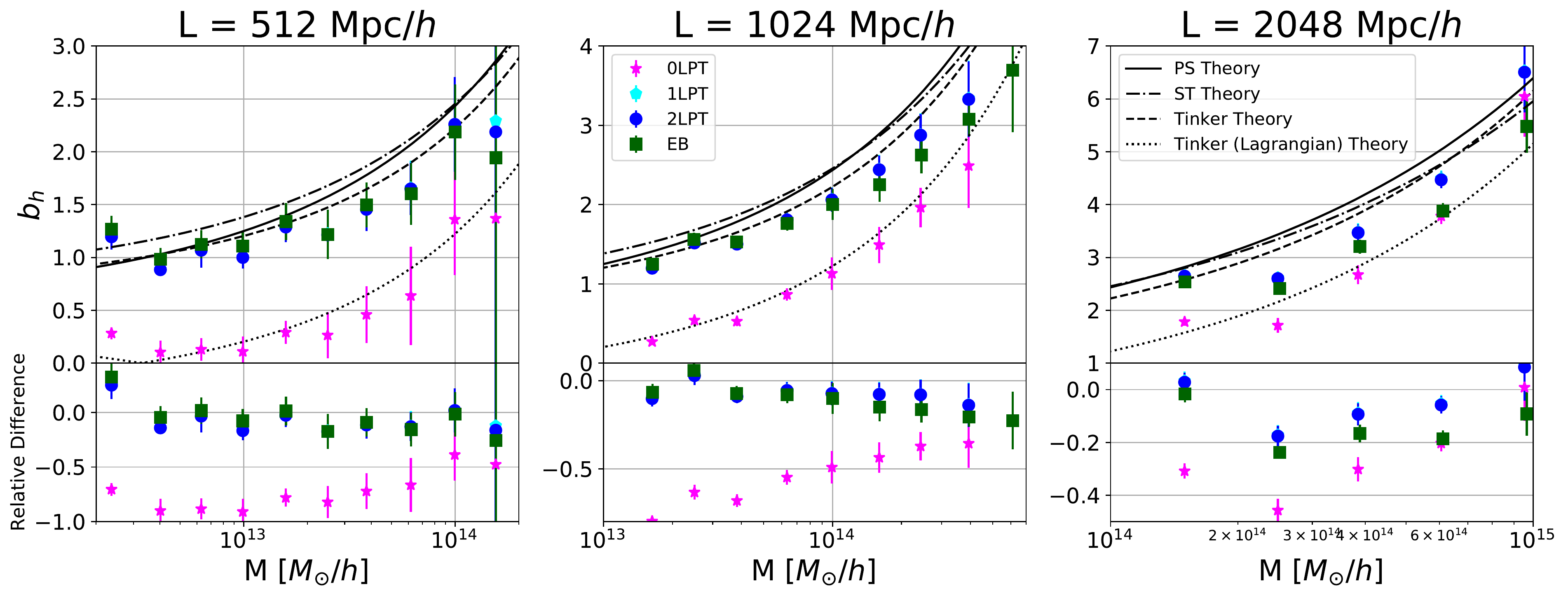}
	\caption{Mean linear halo bias for the eight realizations of each of the three box sizes generated using the SB barrier, with 0LPT ({\it pinks stars}), 1LPT ({\it cyan pentagons}) and 2LPT ({\it blue dots}) and using the EB barrier (green squares), along with $\pm 1\sigma$ errors. Also show are predictions of \citep{Press} (PS, solid line),  \citep{Sheth2} (ST, dotted-dashed line), \citep{Tinker2} (Tinker, dashed line) and the Lagrangian version of the Tinker's fit (dotted line). The bottom panels show the relative differences of each catalogue with respect to Tinker's fit. The panels show results for boxes with sizes of  $L=512$, $1024$ and $2048$ Mpc$/h$ from left to right.}
	\label{fig:Bias_test}
	\end{center}
	\end{figure*}

    In Fig.~\ref{fig:Bias_test} we present the linear halo bias with the same labels of the previous figure. We compute the linear bias $b_h$ using the fact that on large scales it should be scale independent and given by 
\begin{equation}
b_{h}(M) = \sqrt{\frac{P_{hh}(k|M)}{P_{mm}(k)}} \,,
\label{eq:Halo-Halo_bias}
\end{equation}
where $P_{hh}(k|M)$ is the halo-halo power spectrum for halos with mass $M$ and $P_{mm}(k)$ is the theoretical linear matter power spectrum computed using \texttt{CAMB} \citep{CAMB}.

	For each bin of halo mass, we measure the halo-halo power spectrum and fit the ratio with the linear matter power spectrum using a linear function in $k$ up to $k=0.1$  $h/$Mpc. We then take the constant term as the linear halo bias. The mass bins are chosen as logarithmically spaced and only bins with more than $750$ halos in each realization ($6000$ halos in total) were considered.

    We can see in Fig.~\ref{fig:Bias_test} that the results from the two barriers are very similar and very close to theoretical predictions. In particular, there is an agreement within $20\%$ with Tinker's prediction (dashed line). Our method  underestimates the linear halo bias for the most massive halos, in comparison with theory predictions, but these points are evaluated using a small number of halos, and shot noise dominates the estimate of $P_{hh}(k|M)$. 
    In principle these results could be improved by using other functional forms for the scale-dependent bias and detailed modelling of non-Poissonian shot noise, but these extra corrections are out of the scope of this work.
    
    The linear bias for the two different barriers display the same relative behaviour as the abundance and the theoretical predictions, where the difference between the two changes sign at $ M \approx 10^{14} M_{\odot}/h$. However, note that the linear bias is larger for the EB catalogues at low masses.
    
    As expected by perturbation theory, the 2LPT and 1LPT catalogues have the same linear bias. This happens because the 1LPT theory already provides the correct linear correction to halos. Higher orders in LPT will be relevant for the quasi-linear scales and for correlation functions of more than 2 points.
    
    In Fig.~\ref{fig:Bias_test} we also plot the Lagrangian version of Tinker's formula in dotted line. The Lagrangian bias is given by $b_{h}^{(L)} = b_{h} - 1 $, where $b_{h}^{(L)}$ is the linear halo bias in  Lagrangian space. We see that the 0LPT catalogues, where LPT is not applied, has a linear bias well described by the Lagrangian bias prediction. This shows that the displacement of the halos using LPT provides the correct mapping between Lagrangian and Eulerian spaces.
    
    In conjunction, Figs.~\ref{fig:Abundance_test} and \ref{fig:Bias_test} indicate that our method produces halo catalogues with an agreement of $25 \%$ relative to Tinker's fitting formulae. This agreement shows that the halo finding procedure (step 2) gives the correct halo abundance and that the displacement of these halos using LPT (step 3) puts them in the correct positions, at linear scales.
    
    In \S~\ref{sec:Comparisons_Simulations}
we compare some of our results to full N-body simulations. An investigation of the $k$-dependence in the power spectra and bispectra is performed in Appendix~\ref{app:Variation}.
    
\subsection{Computational Requirements}

    In order to check the computation time and memory required by the current implementation of our method, we ran our code with different values of box size and different numbers of grid cells. In all tests we fixed the cosmology to that of the N-body simulation given in the next section, the SB model and outputted only halos with more than $8$ particles. We also ran the code using a single processor, even though our current implementation allows for \texttt{OPENMP} parallelization. 
    
\begin{figure}
\begin{center}
\includegraphics[width=\linewidth]{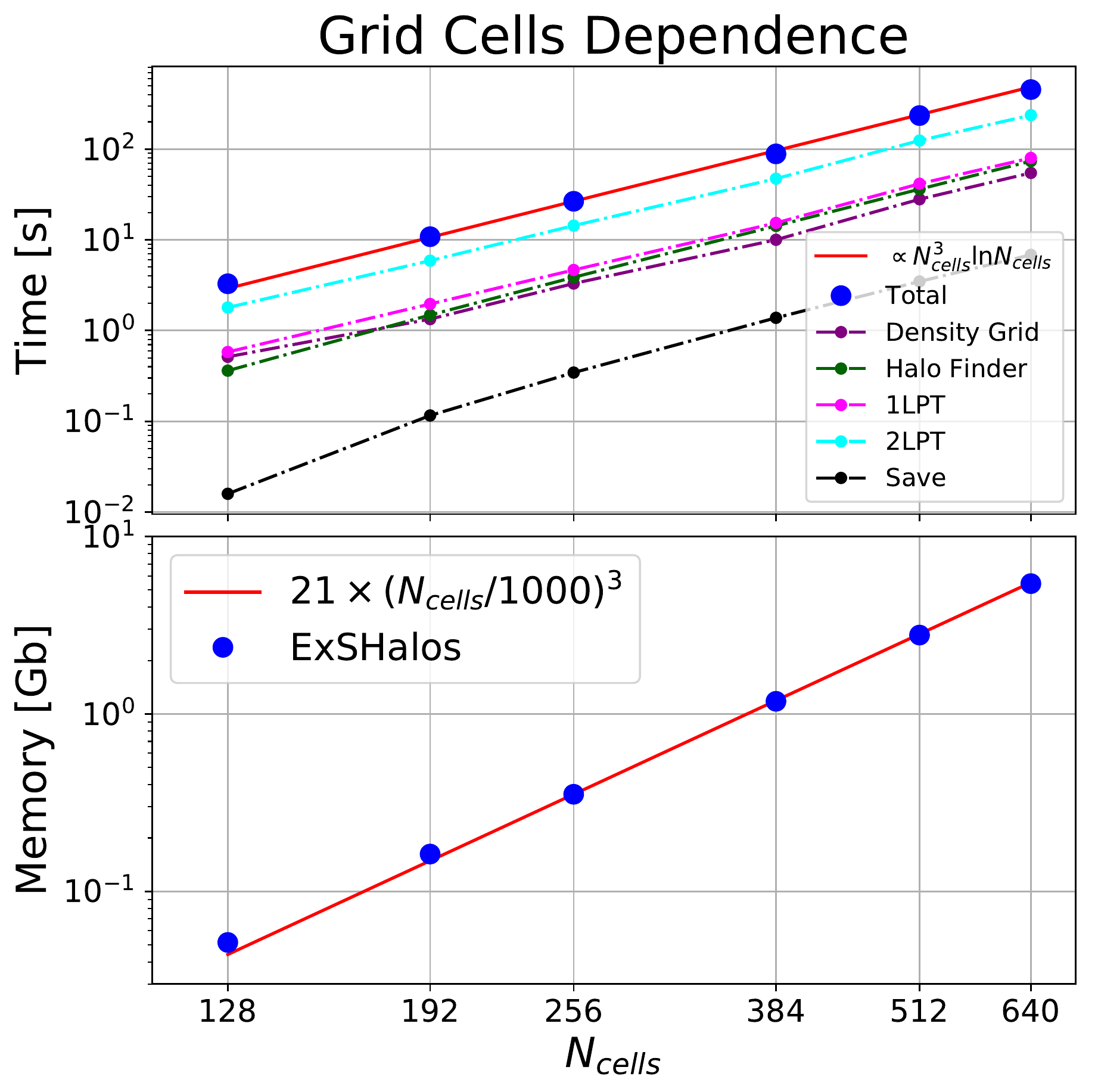}
\caption{Computation time ({\it upper panel}) and memory ({\it lower panel}) required to run the current implementation of our method as a function of the number of grid cells per dimension in the box. We used 
a box with size $L=1024$~Mpc$/h$ in each direction and the SB model. 
Blue dots show the total requirements of our code and the red line is a power law proportional to $N_{cells}^{3}$. 
The dashed-dotted lines with points show the partial times needed in each step of the code: generation of the initial density grid (purple), halo finding (green), displacement of halos with 1LPT (pink), displacement of halos with 2LPT (cyan) and the time to save in memory all halos with at least eight particles.}
\label{fig:Nc_dep}
\end{center}
\end{figure}

    In Fig.~\ref{fig:Nc_dep} we show the dependence of the computation time and memory needed to generate a box with size $L=1024$ Mpc$/h$ as function of the number of grid cells per dimension. These results are also presented in Table~\ref{tab:Performace}, and reflect the computation time for a single 4 GHz processor.

\begin{table}
\centering
\caption{Number of grid cells per dimension (resolution), total computation time and maximum memory needed by our method code to generate halos in a box with size $L=1024$~Mpc$/h$ in each direction. We also show the expected memory given by Eq.~\eqref{eq:Memory}.
These results are also shown in Fig.~\ref{fig:Nc_dep} and reflect the computation time for a single 4 GHz processor.
}
\begin{tabular}{c c c c }
\hline
$N_{cells}$ & Time [$s$] &  Memory [Gb] &  Eq.~\eqref{eq:Memory} [Gb]\\
\hline
$128$ & $3.28$ & $0.051$ & $0.044$ \\
$192$ & $10.83$ & $0.162$ & $0.147$ \\
$256$ & $26.62$ & $0.352$ & $0.352$ \\
$384$ & $88.64$ & $1.178$ & $1.189$ \\
$512$ & $234.52$ & $2.782$ & $2.819$ \\
$640$ & $453.25$ & $5.427$ & $5.505$ \\
\hline
\label{tab:Performace}
\end{tabular} 
\end{table}

    Our current implementation requires five float numbers ($4$ bytes numbers) per grid cell to store the density field, in real and Fourier spaces, as well as the three components of the displacement field. The code also requires slightly more memory to store information about the halos, but it is a subdominant contribution. We propose the following parametrization for the memory required:
    \begin{equation}
    \mbox{Memory} \approx 21 \times \left( \frac{N_{cells}}{1000} \right) ^{3} \mbox{\texttt{Gb}} \,,
    \label{eq:Memory}
    \end{equation}
    where $N_{cells}$ is the number of grid cells per dimension.
    
    This parametrization slightly underpredicts the memory needed for small numbers of grid cells, $N_{cells} < 256$, and slightly overpredicts the memory needed for large numbers of grid cells. This also works well for the range of box sizes considered here, $128 - 4096$ Mpc$/h$.
    
    In the lower panel of Fig.~\ref{fig:Nc_dep}, we compare our parametrization with results coming from direct measurement of the memory usage. It becomes clear that our parametrization \eqref{eq:Memory} is reasonable and that the memory required scales almost linearly with the number of grid cells per dimension, as expected. 
    
    In the upper panel of Fig.~\ref{fig:Nc_dep}, we can see how each step contributes to the total computation time. The time to compute the 2LPT displacement, which requires two fast Fourier Transforms (FFT's), dominates the computation time. The times to construct the density grid, perform the 1LPT displacement and find the halos are of the same order. Finally, the time to save the halo information is sub-dominant. This last time is strongly dependent on the minimum number of particles considered to save a halo.
    
    As the main contribution to the computation time comes from the 2LPT displacement, the total time is expected to be of order $\mathcal{O}(N_{cells}^{3} \ln{N_{cells}})$ because of the FFT's. This is verified in Fig.~\ref{fig:Nc_dep}, since for large values of $N_{cells}$ the blue dots are slightly above the red line, which scales as $N_{cells}^{3}$.
	
\begin{figure}
\begin{center}
\includegraphics[width=\linewidth]{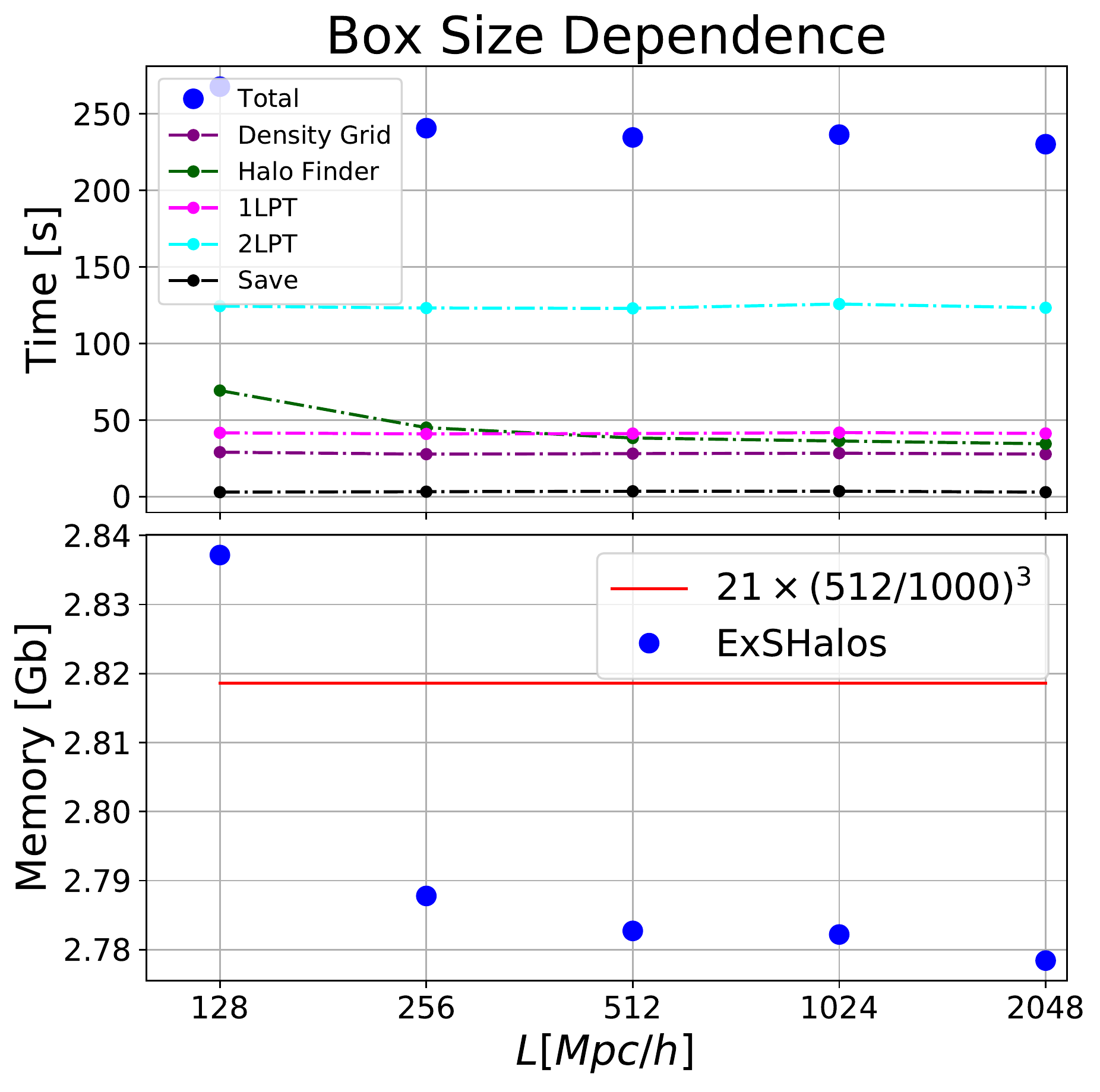}
\caption{The same as in Fig.~\ref{fig:Nc_dep} but as function of box size for a resolution of $512$ grid cells per dimension.}
\label{fig:L_dep}
\end{center}
\end{figure}

    In Fig.~\ref{fig:L_dep} we present the computation time and memory needed to generate halo catalogues with different box sizes. In this comparison we fixed $N_{cells} = 512$ and used the same cosmology, barrier and minimum number of particles as in Fig.~\ref{fig:Nc_dep}. 
    
    In the lower panel of Fig.~\ref{fig:L_dep}, we see that the memory needed changes by less than $1.5 \%$ from the smallest to the largest box size considered. This shows that the main contribution to the memory requirement comes from quantities allocated in grid cells.
    
    In the upper panel of Fig.~\ref{fig:L_dep}, we see that the computation time changes more significantly, around $15 \%$ for the smalest box size considered. This change comes from the time required in the halo finding step, which is larger for smaller boxes with more halos. However, the computation time and memory needed almost do not change for simulations with resolution in the range $0.5 - 4.0$ Mpc$/h$, values typically used.

\section{Comparison with N-Body Simulations}
\label{sec:Comparisons_Simulations}





	In this section we compare our halo catalogues with halo catalogues measured from N-body simulations, where halos are found using spherical overdensity (SO) and friends-of-friends (FoF) methods. We compare the mass function, the linear halo bias, the two- and three-point density functions and the two-point function of the velocity divergence field.
	
\subsection{Halo Catalogues}
\label{sec:Halo_Catalogues}

	We now compare our halo catalogues with 
	the SO and FoF halos found in the MultiDark simulation \citep{Prada}. 
    In all catalogues we use the same cosmology of MultiDark, namely: $h = 0.7$, $\Omega _{m} = 0.27$, $\Omega _{\Lambda} = 0.73$, $\Omega _{b} = 0.0469$, $n_{s} = 0.95$ and $\sigma _{8} = 0.82$.
    
    The halo catalogues used for our comparisons are described in Table~\ref{tab:Catalogues}, which shows the catalogue's name, the box length of each simulation, the number of particles (for N-body simulations) or the number of grid cells (for ExSHalos catalogues), the mass of each particle or the mass within each cell, the number of realizations, and specific characteristics of each catalogue. 
    
\begin{table*}
\centering
\caption{Specifications of the halo catalogues used in the comparisons with the simulations, including their names, box length, number of particles (for the N-body catalogues) or number of grid cells (for the halos generated using our approach), minimum mass resolution (the mass of each particle, or the mass in each grid cell), number of realizations for each catalogue, and some specific characteristic of the catalogue.}
\begin{tabular}{l c c c c r}
\hline
Name & $L$ [Mpc$/h$] & $N_{p}/N_{d}$ & $M_{\rm min}$ [$M_{\odot}/h$] & Realizations &  Characteristics \\
\hline
BDMV & $1000$ & $2048 ^{3}$ & $8.721\times 10^{9}$ & 1 & Spherical Overdensity with $\Delta = 360 \bar{\rho} _{m}$  \\
FoFc & $1000$ & $2048 ^{3}$ & $8.721\times 10^{9}$ & 1 & Friends-of-Friends with $l = 0.2$  \\
\hline
SB & $512$ & $512 ^{3}$ & $7.491\times 10^{10}$ & 8 & Static Barrier with 2LPT \\
EB & $512$ & $512 ^{3}$ & $7.491\times 10^{10}$ & 8 & Ellipsoidal Barrier with 2LPT \\
\hline
\label{tab:Catalogues}
\end{tabular} 
\end{table*}

	As shown in Table~\ref{tab:Catalogues}, we used four different halo catalogues in the comparison: BDMV, where halos were found in the MDR1 MultiDark simulation using the SO technique with an overdensity $\Delta=360$ (close to the virial overdensity in this cosmology);  FoFc catalogues, which are also based on the MDR1 simulation, but uses the FoF algorithm with a linking length of 0.2; SB, which employs our method with a static barrier and 2LPT displacement; and EB, which employs our method with an ellipsoidal barrier and 2LPT displacement.
    
    We consider both the SO and FoF halos from the simulations, since SO halos should be closer to the excursion set predictions, while FoF halos are  most commonly used in the literature. We expect that the SO halos (BDMV) are more similar to the halos generated with our method, which exploits ideas from the excursion set theory. However, it should also be possible to fit for barrier parameters in order to recover the FoF halo properties. This is similar to what happens in the theoretical predictions of the halo mass function and linear bias, where analytical predictions are computed for SO halos, but not for FoF halos \citep{Maggiore1, Achitouv1, Achitouv2}. 
    These two different simulated halo catalogues also give an estimate of the uncertainty in the halo properties due to differences in halo definitions. 

	In our comparisons we divided the simulated catalogues into eight sub-catalogues, with box lengths of $500$ Mpc$/h$ and with $1024 ^{3}$ particles on average, in order to compute standard deviations for each observable. Although the simulated catalogues have eight times more mass resolution than our catalogues, we will see that the halos generated with our method are in good agreement with theory predictions at small masses.
	
	Another important aspect to note is that the two simulated catalogues (BDMV and FOFc) have the same initial conditions, i.e. they define different halos found with two techniques applied to the same particle catalogue. Likewise, the two catalogues generated by our method (SB and EB) also have the same initial conditions, which however differ from the simulated one. Therefore it is possible to compare, point by point, the results of the N-point functions between the two simulated catalogues and the two ExSHalos catalogues.
	
	The parameters of the ellipsoidal barrier used in our EB catalogues were fitted comparing the power spectrum of the EB halos with that from the BDMV catalogues, using a Gaussian weight in the $\chi ^{2}$ to cut off non-linear scales, since we require agreement with simulations only for linear and mildly non-linear scales.

\subsection{Mass Function and Linear Bias}
\label{sec:Abundance_Bias}

\begin{figure}
\begin{center}
\includegraphics[width=\linewidth]{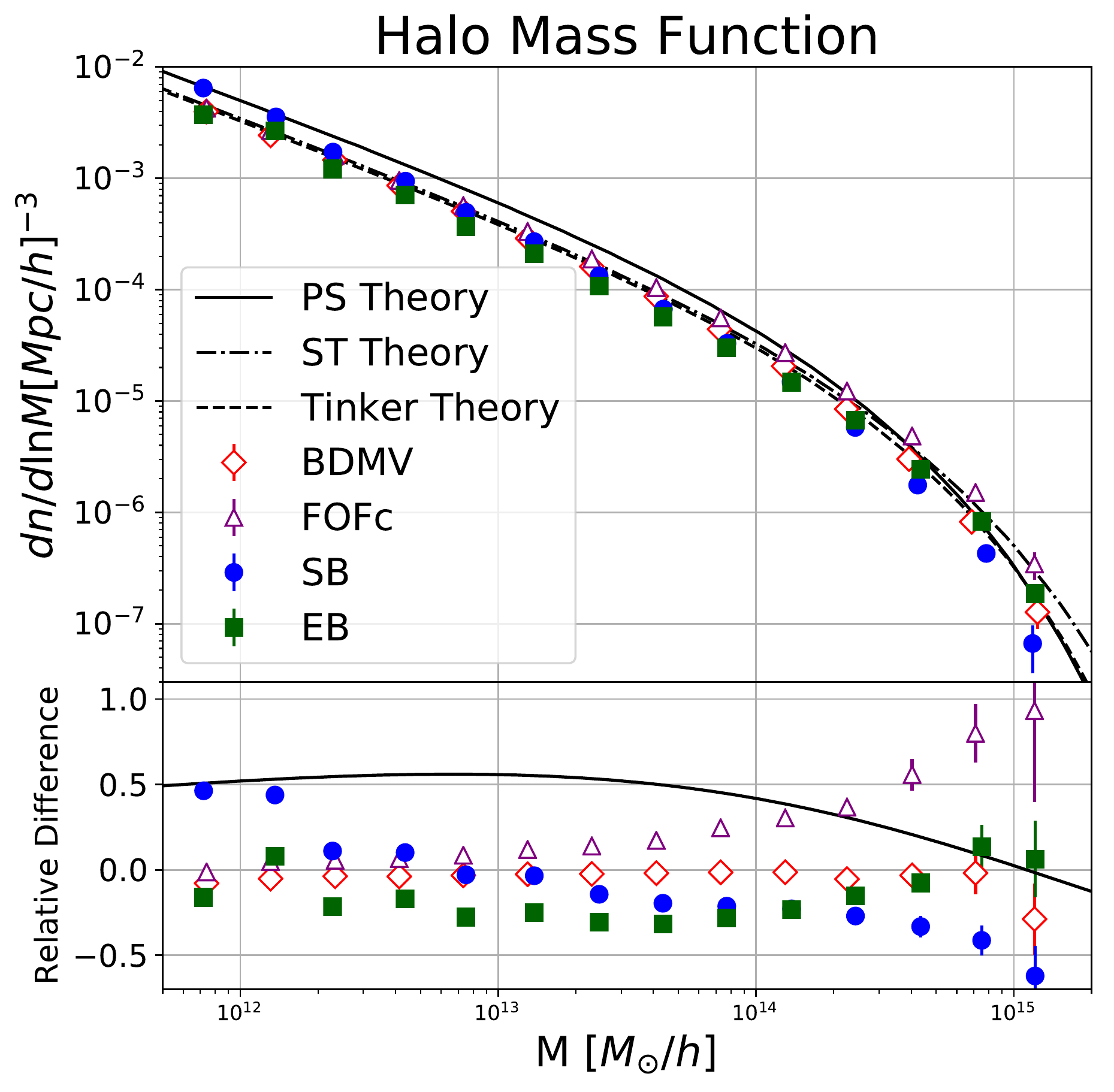}
\caption{Mean halo abundance for the eight realizations of each halo catalogue described in Table~\ref{tab:Catalogues}: SB (blue dots), EB (green squares), BDMV (red diamonds) and FoFc (purple triangles), with $\pm 1\sigma$ error bars. We also show predictions from \citep{Press} (solid line), \citep{ST} (dotted-dashed line) and \citep{Tinker2} (dashed line). The bottom panel shows the relative differences of each case with respect to Tinker's fit.}
\label{fig:Abundance_NBody}
\end{center}
\end{figure}

	In Fig.~\ref{fig:Abundance_NBody} we show the halo abundance for the four different catalogues of Table~\ref{tab:Catalogues}. We use all halos with 64 particles or more to compute the abundance for the simulations (BDMV and FoFc) and all halos with 8 particles or more for the halos generated with ExSHalos (EB and SB). We can see that our method generates halo catalogues whose abundance is accurate in this range of masses, even though they have eight times lower resolution compared with simulations. As in Fig.~\ref{fig:Abundance_test}, we also show the theoretical predictions of \cite{Press}, \cite{Sheth2} and \cite{Tinker}.
	
    The BDMV catalogue agrees well with the fit of \cite{Tinker} (within $5\%$), while the FoFc catalogue differs significantly (over $50\%$ at the largest masses). This simply reflects the fact that the halo geometry and mass definitions impact the properties and statistics of the halo catalogue. Both our catalogues (SB and EB) agree with theory predictions (within $30\%$), indicating good accuracy in the halo mass function. 
    
    In particular, the measured abundance of our SB catalogue is in better agreement with the Tinker halo mass function than with the Press-Schechter prediction (solid line). As mentioned earlier, this occurs because the method takes into account the non-Markovian corrections at all orders. Moreover, we are using a cubic top-hat as the window function on the density map, and this is different from the spherical top-hat function used in theoretical predictions of the mass function and linear bias. These two differences from the excursion set theory make it difficult to predict exactly the final mass function and linear bias from our method.  

\begin{figure}
\begin{center}
\includegraphics[width=\linewidth]{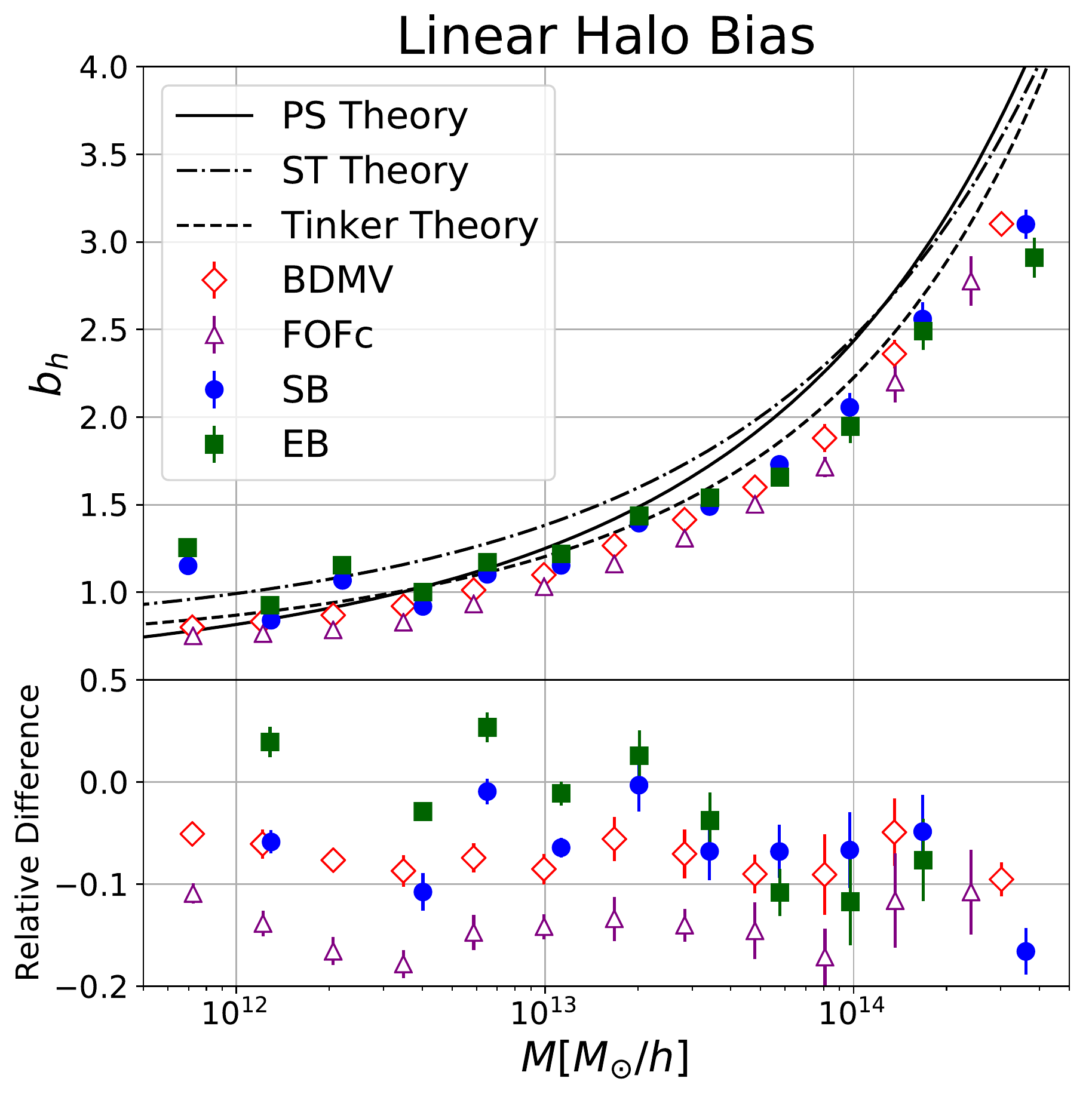}
\caption{Mean linear halo bias for the eight realizations of each halo catalogue described in Table~\ref{tab:Catalogues}: SB (blue dots), EB (green squares), BDMV (red diamonds) and FoFc (purple triangles), with $\pm 1\sigma$ errors. We also show predictions from \citep{Press} (solid line), \citep{ST} (dotted-dashed line) and \citep{Tinker2} (dashed line). The bottom panel shows the relative differences of each catalogue with respect to the halo bias from \citep{Tinker2}.}
\label{fig:Bias_NBody}
\end{center}
\end{figure}

	In Fig.~\ref{fig:Bias_NBody}, we present the linear halo bias for the four halo catalogues listed in Table~\ref{tab:Catalogues}, along with the same three theoretical predictions used for the mass function. Here we compute the linear bias in the same way as in the previous section, i.e. 
	as the ratio between the measured density-density halo power spectrum in some mass bin and the theoretical linear density-density matter power spectrum given by \texttt{CAMB}. We fit a first order polynomial using all points with $k < 0.1$~Mpc$/h$, and take the constant term as the linear bias. 

	As shown in Fig.~\ref{fig:Bias_NBody}, the linear bias of halos generated by our method agrees with BDMV's linear bias within $1 \sigma$, and agrees with the bias prediction of \cite{Tinker2} within $10\%$ over all mass scales. Biases coming from different barrier models do not display large differences, but the bias coming from the FOFc catalogue is lower than the bias of the BDMV catalogue for all masses. This relative difference between the linear bias of the two N-body catalogues will be also apparent in the correlation functions.

\subsection{Density Two-Point Function}
\label{sec:Power_Spectrum}

\begin{figure}
\begin{center}
\includegraphics[width=\linewidth]{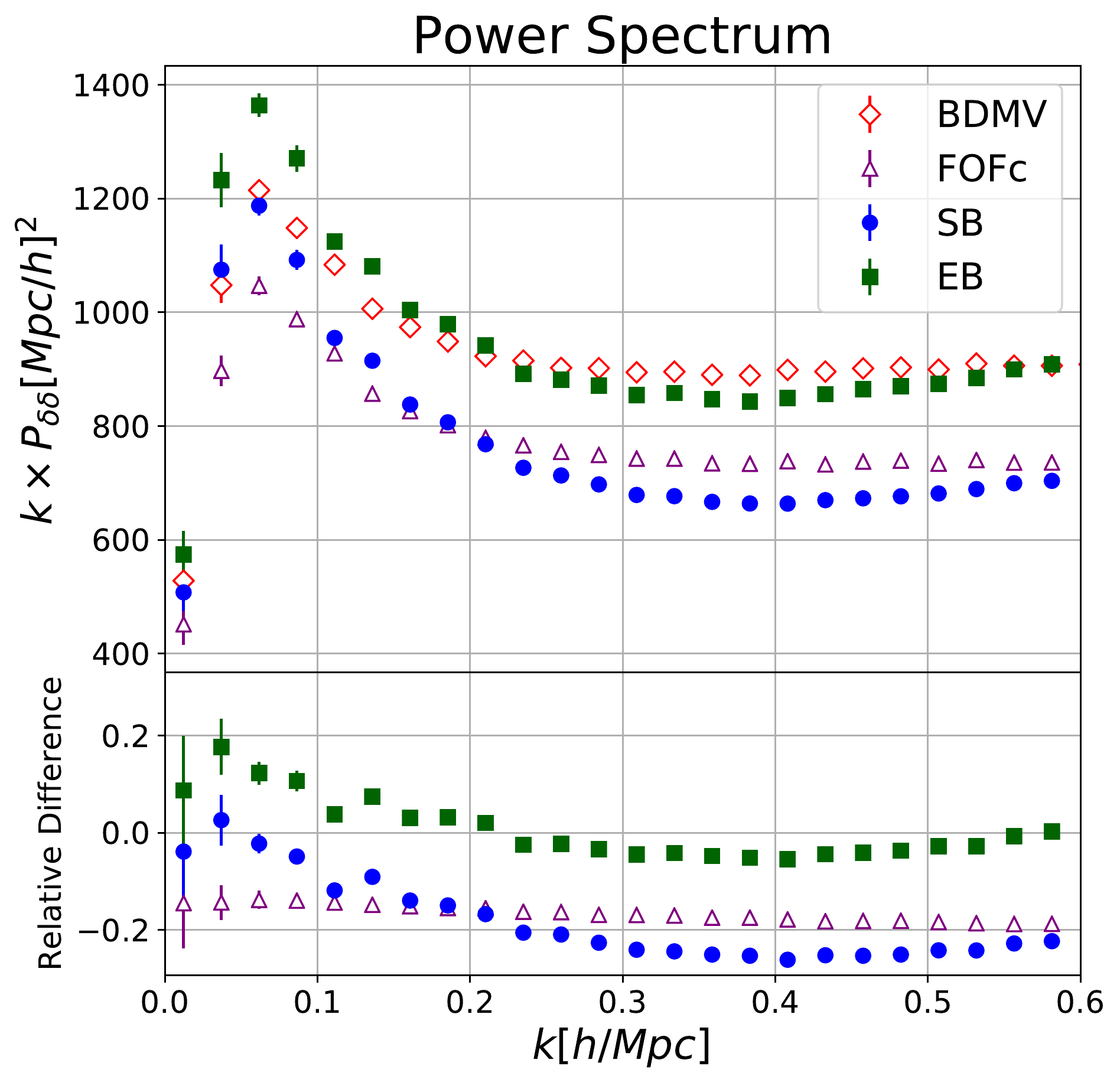}
\caption{Mean density-density halo power spectrum for the eight realizations of each halo catalogue described in Table~\ref{tab:Catalogues}: SB (blue dots), EB (green squares), BDMV (red diamonds) and FoFc (purple triangles), with $\pm 1\sigma$ errors. The bottom panel shows the relative differences of each catalogue with respect to the BDMV power spectrum.}
\label{fig:Phalos_NBody}
\end{center}
\end{figure}

	In Fig.~\ref{fig:Phalos_NBody}, we show the density-density halo power spectrum for the four halo catalogues considered in this section. We measured the power spectrum considering all halos with $M \geq 5\times 10^{12} M_{\odot}/h$, mass range where the abundance agrees with the fit from \cite{Tinker}. We also chose a large mass range in order to reduce the shot noise contribution in the measurement. We used $256^{3}$ cells in our grid to compute the power spectrum, which is sufficient to probe linear and quasi-linear scales.

	In the lower panel of Fig.~\ref{fig:Phalos_NBody}, we can see that our catalogues generated using the SB barrier have $\approx 10 \%$ difference with respect to the BDMV catalogue for $k \leqslant 0.15$ $h/$Mpc, and underestimate the power spectrum for larger values of $k$. We have an agreement between our catalogues and simulations for linear scales, and a controlled failure for small scales, where we find a lack of power. This is the expected behaviour for the density-density halo power spectrum from the halo model, where non-linear scales will be corrected by the appropriate halo density profile, evading a double counting of power.
    
    Moreover, we can see the effect of different halo definitions in the N-body simulation (BDMV and FoFc catalogues), and different barrier choices (SB and EB catalogues). In the lower panel of Fig.~\ref{fig:Phalos_NBody}, it becomes clear that the power spectra for the halos measured from the simulation differ almost by a constant factor. This shows that the $k$-dependence of the spectrum changes weakly with the specific choices of the halo finder, but these choices strongly affect the linear bias and abundance, as shown in the last subsection. In the case of halos generated by our method we can see a similar behaviour, where the SB and EB spectra also differ by a factor nearly constant \footnote{This constant difference becomes clear in the first row of Fig.~\ref{fig:Powers_Barriers}, where we compare the power spectra measured for the two barrier choices.}.  
    
    Now looking at the spectrum of our EB catalogue, we can see that the first points (large scales) are slightly above those from the simulation, while the mildly non-linear scales have a good agreement with the BDMV catalogues. This happens because we fitted the parameters of the ellipsoidal barrier Eq.~\eqref{eq:Barriers} to agree with the simulation up to these mildly non-linear scales ($k \approx 0.3$ Mpc$/h$). Since the power spectrum on small scales has small error bars, the fit is dominated by these scales. This effect could in principle be corrected by giving higher weights for large scales.  In any case, our main point here is to show that one can improve the agreement with simulations by fitting parameters of the ellipsoidal barrier. 
    
    We also note that by considering these free parameters in the barrier, our method becomes very similar to the predictive class where we find halos directly in the initial conditions, and then fit some parameters in order to provide better agreement with simulations. 

\subsection{Density Three-Point Function}
\label{sec:Bispectrum}

\begin{figure*}
\begin{center}
\includegraphics[width=0.32\linewidth]{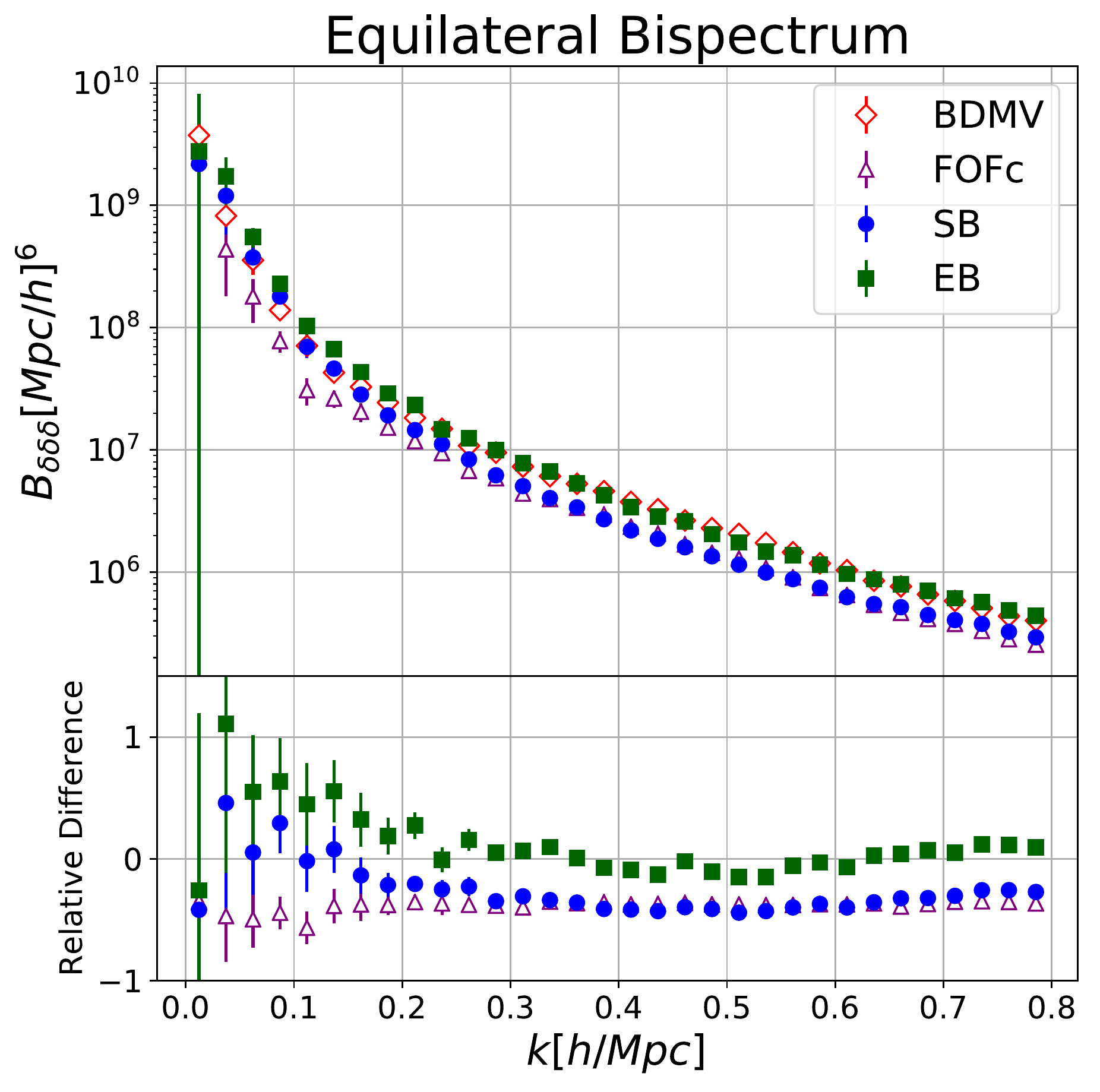}
\includegraphics[width=0.32\linewidth]{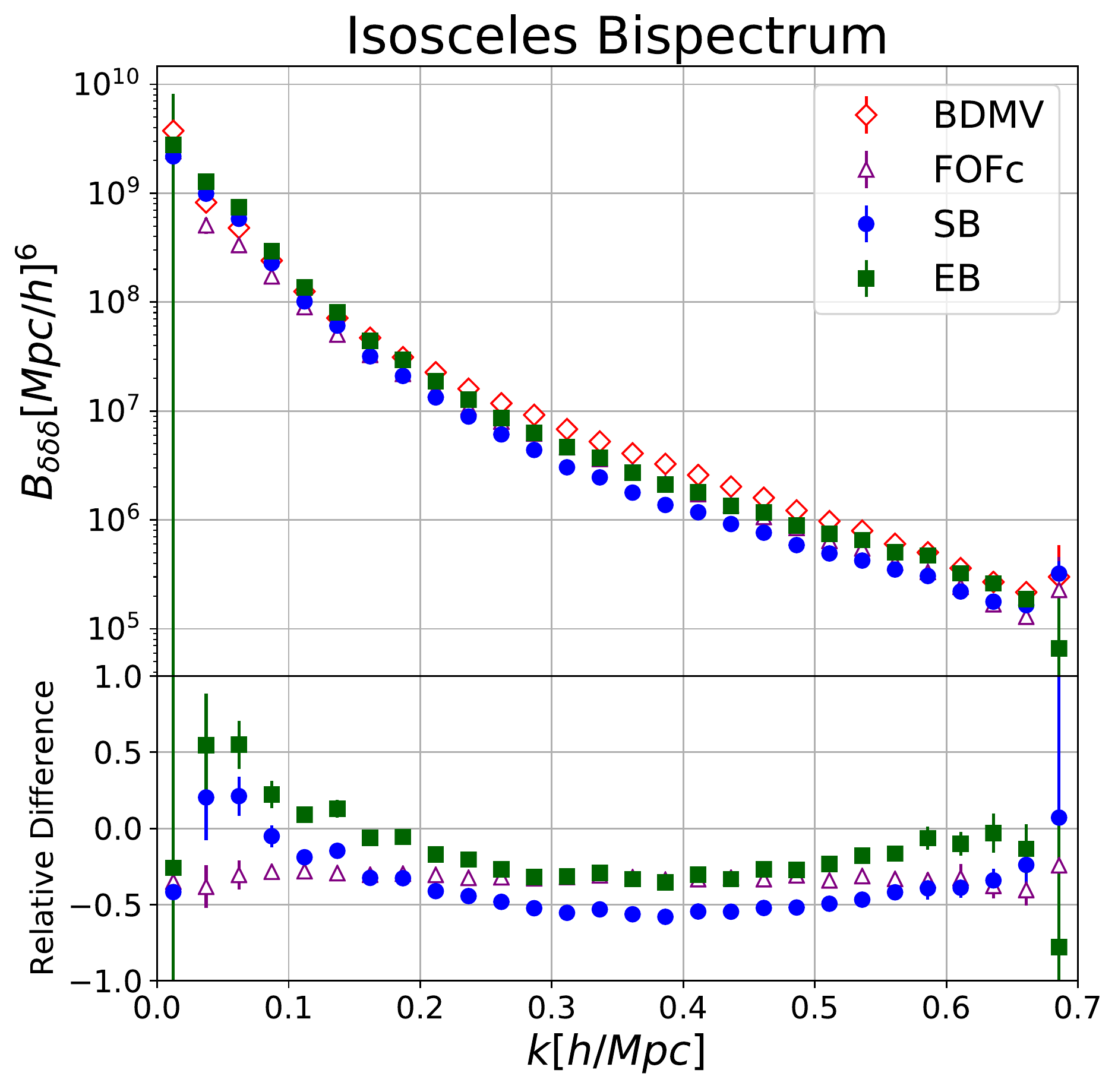}
\includegraphics[width=0.32\linewidth]{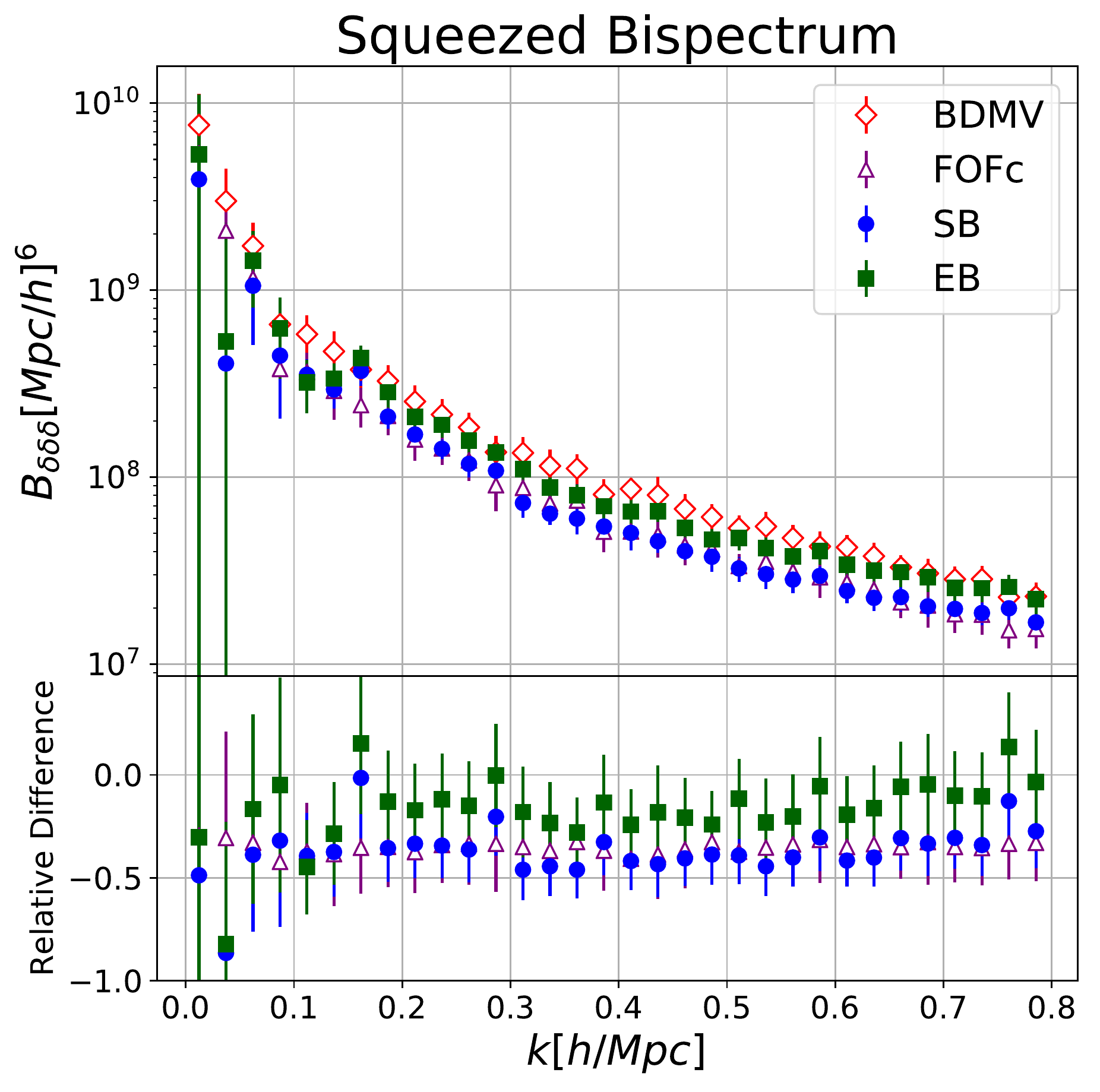}
\caption{Mean density-density-density halo bispectrum for the eight realizations of each halo catalogue described in Table~\ref{tab:Catalogues}: SB (blue dots), EB (green squares), BDMV (red diamonds) and FoFc (purple triangles), with $\pm 1\sigma$ errors. The bottom panels show the relative differences of each catalogue with respect to the BDMV bispectrum. The first column presents the bispectra for equilateral triangles, the second for isosceles triangles and the third for the squeezed limit.}
\label{fig:Bhalos_NBody}
\end{center}
\end{figure*}

	In Fig.~\ref{fig:Bhalos_NBody}, we present bispectrum measurements for the same four halo catalogues. As in the case of the power spectrum, we gain we use all halos with $M \geq 5\times 10^{12} M_{\odot}/h$. However, now we use only $128^{3}$ grid cells, because the measurement of the bispectrum is considerably slower than that of the power spectrum. We measure the bispectrum using the fast method described in \cite{Watkinson}, where we do not explicitly compute the average over all triangles, but instead compute six more fast Fourier transforms. We compute the bispectrum for three different triangular configurations: equilateral triangles ($k_{1} = k_{2} = k_{3} = k$), isosceles triangles ($k_{3} = 2 k_{2} = 2 k_{1} = 2k$) and the squeezed limit ($k_{1} = k_{2} = k$ and $k_{3} = 2 \pi/L$).

	The bispectrum of the SB catalogue agrees with the BDMV measurement for $k<0.15$ Mpc$/h$ in the equilateral configuration, and for $k<0.1$ Mpc$/h$ in the isosceles configuration.  In the squeezed limit the SB bispectrum is slightly below the BDMV bispectrum ($\approx 30 \%$) but the error bars are  larger in this case. These results are of same order than the results for the power spectrum. The catalogues generated by our method also recover the bispectrum at linear scales and have a lack of power at non-linear scales.
    
    Another feature that also appears in the bispectrum is the constant multiplicative difference when comparing the BDMV and FoFc catalogues, or when comparing the SB and EB catalogues\footnote{Again this constant difference between the SB and EB bispectra becomes clear in Fig.~\ref{fig:Bis_Barries}, where the two are directly compared.}.
    
    The behaviour of the bispectrum for the EB catalogues is similar to that of the power spectrum, where we have a slight overestimation at linear scales, and a better agreement at non-linear scales in comparison with the SB results. It is important to note that we use only the power spectrum measurements to fit the parameters of the ellipsoidal barrier, but the bispectrum measurements are also improved.

\subsection{Velocity Two-Point Functions}
\label{sec:Vel_Power}

\begin{figure}
\begin{center}
\includegraphics[width=\linewidth]{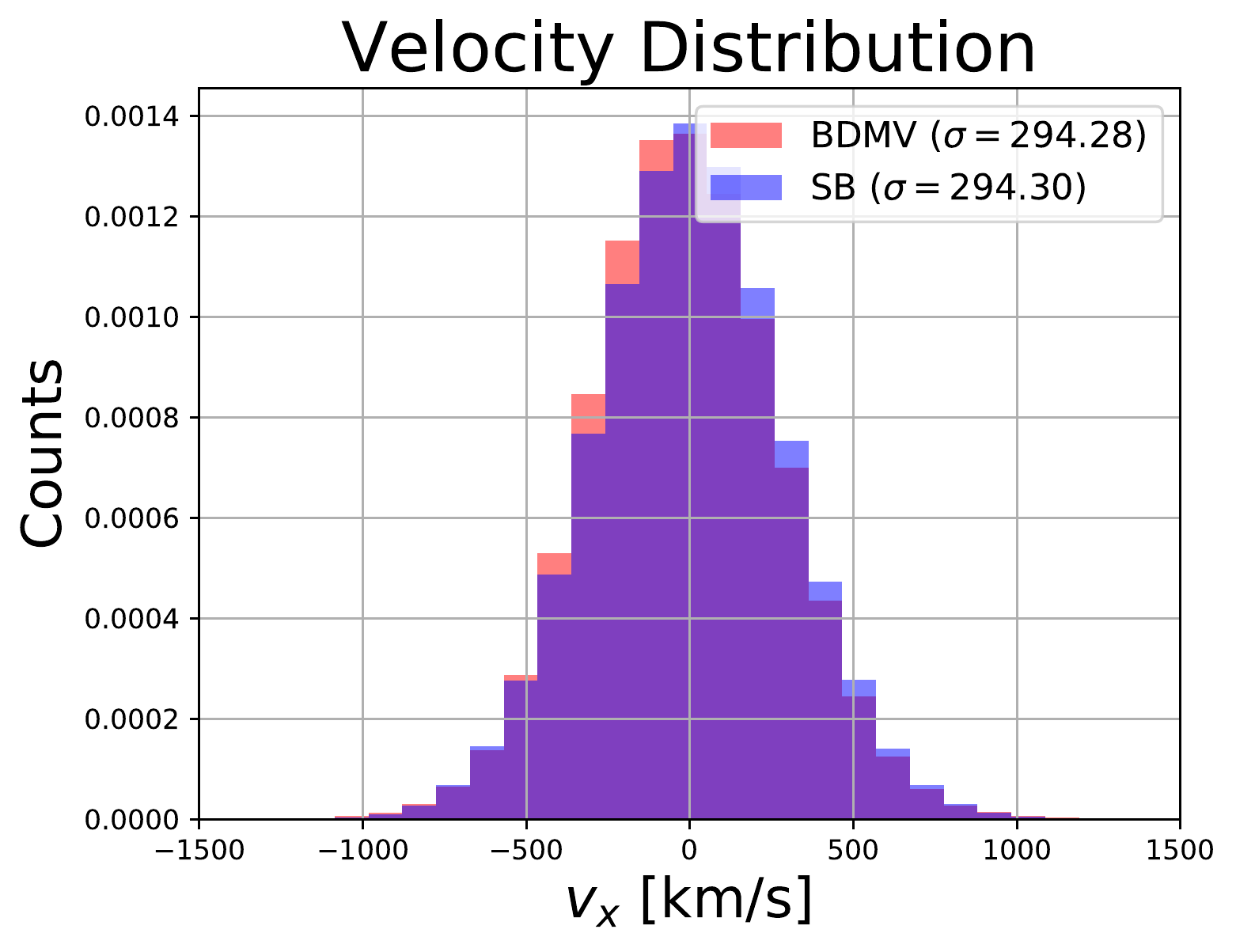}
\caption{Normalized distribution of the $x$ component of the halo velocity for the BDMV halos (in red) and SB halos (in blue), along with standard deviations for each catalogue. 
}
\label{fig:Vel_dist}
\end{center}
\end{figure}

    In Fig.~\ref{fig:Vel_dist}, we show the distribution of the $x$ component of the velocity for the BDMV and SB halos (the other components give similar results). Our method reproduces very well the velocity distribution with a negligible difference in the standard deviation ($<0.007 \%$). To compute these distributions we used all halos with $M \geq 5 \times 10^{12} M_{\odot}/h$ of one realization, although the other realizations show similar results. Therefore, the 2LPT displacement is sufficient to give the correct velocity distribution to the halos.

\begin{figure*}
\begin{center}
\includegraphics[width=0.49\linewidth]{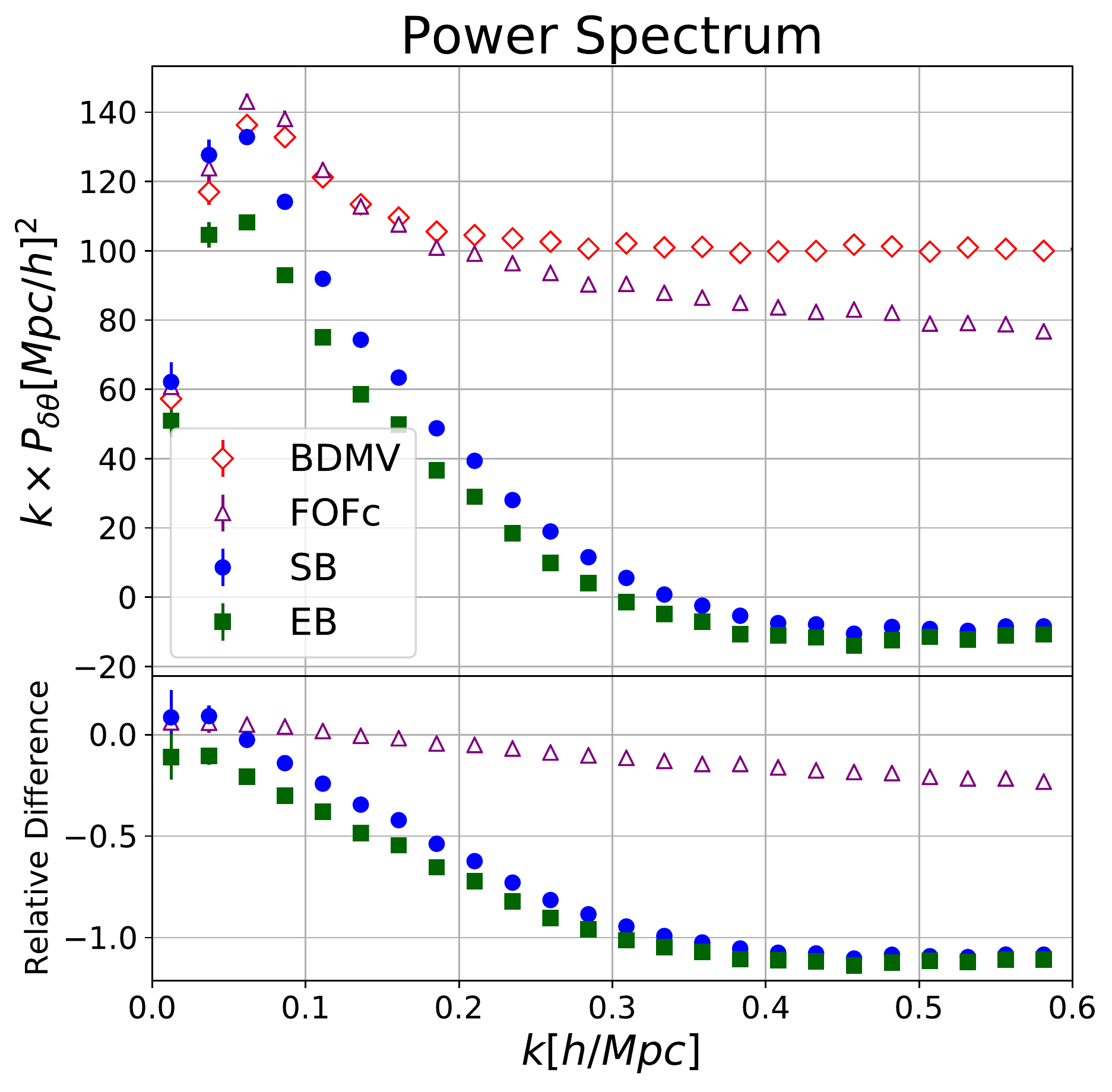}
\includegraphics[width=0.49\linewidth]{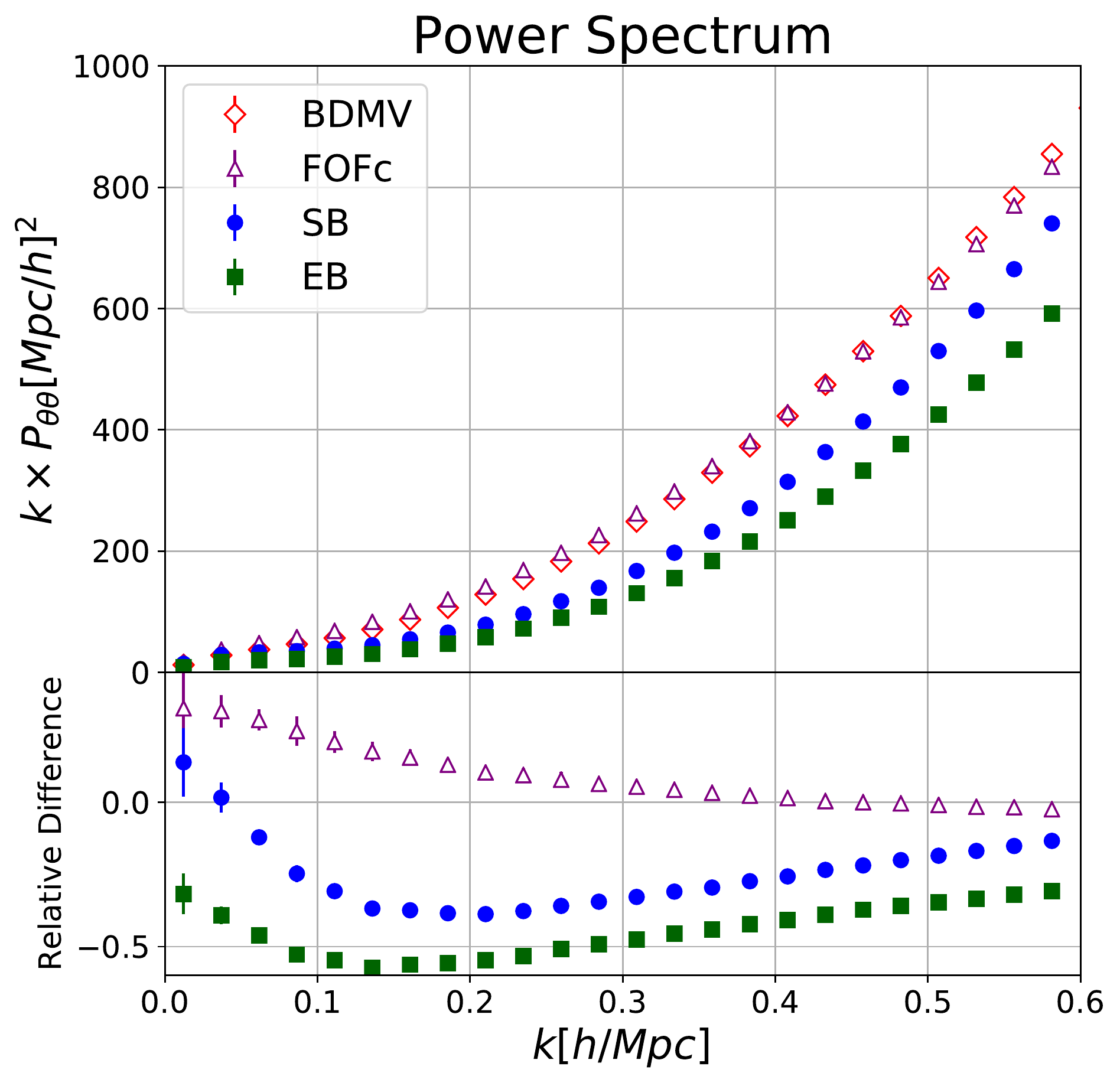}
\caption{Mean density-velocity ({\it left panel}) and velocity-velocity ({\it right panel}) halo power spectrum for the eight realizations of each halo catalogue described in Table~\ref{tab:Catalogues}: SB (blue dots), EB (green squares), BDMV (red diamonds) and FoFc (purple triangles), with $\pm 1\sigma$ errors. The bottom panels show the relative differences of each catalogue with respect to the BDMV power spectrum.}
\label{fig:Ptthalos_NBody}
\end{center}
\end{figure*}

	In order to consider velocity and density correlations, in Fig.~\ref{fig:Ptthalos_NBody} we present the density-velocity and the velocity-velocity halo power spectra for the four halo catalogues. Again we take all halos with $M \geq 5 \times 10^{12} M_{\odot}/h$, and we use $256^{3}$ grid cells (as in the density-density power spectrum). In fact, we compute the cross-spectrum $\left\langle \delta \theta \right\rangle$ of the density $\delta$ and the velocity divergence field $\theta$, as well as the auto-spectrum $\left\langle \theta \theta \right\rangle$ of the velocity divergence field. This measurement gives us  information about halo velocities, which would be necessary to eventually construct halo and galaxy catalogues in redshift space.

    In Fig.~\ref{fig:Ptthalos_NBody}, we see that the SB catalogues recover the density-velocity and velocity-velocity power spectrum on the largest scales, but there is a lack of power on small scales when compared to the case of the density-density power spectrum and bispectrum ($k>0.05$ Mpc$/h$). Moreover, the results for the EB catalogues are worse over all scales, also in distinction to what happens in the density case.
    
    Another difference with respect to the density measurements of the last two subsections is that the differences between the density-velocity and velocity-velocity spectra are not almost constant when looking at the  two simulated or the two ExSHalos catalogues. This shows that the halo definition has a more important impact in the velocities correlations than in the positions of halos.

\section{Conclusions}
\label{sec:Conclusions}

In this work we developed a method for the fast generation of halo catalogues based on the excursion set theory. This method generates halo catalogues much faster than N-body simulations ($\sim 10^{4} - 10^{5}$ times faster for the cases studied here), while still maintaining good agreement with the N-body halos ($\sim 20 - 25 \%$ in the two- and three-point functions at linear scales). When combined with prescriptions for populating halos with galaxies, the algorithm makes possible the fast generation of galaxy catalogues, allowing for the quick estimation of covariance matrices.

Our method consists in an explicit realization of the excursion set theory ideas to find halos in the initial Gaussian density map using a barrier threshold for halo formation in linear theory. In the simplest case, the barrier is simply the spherical collapse parameter $\delta _{c}$. We also use Lagrangian Perturbation Theory (LPT) to map these halos from Lagrangian to Eulerian space as an approximation for the gravitational evolution.

The main features of our method are that it is predictive and it is parameter-free. It is predictive in the sense that, using only a few ingredients, we can generate halo catalogues with the correct halo mass function and correlation functions, without the need to use N-body simulations to fit for those parameters. It is parameter-free in the sense that, in its simplest version (the spherical barrier), only cosmological parameters need to be specified.

We also implemented a barrier with three free parameters that take into account the non-spherical nature of the collapse. These three parameters can be set in order to improve the results of the catalogues in the comparison with simulations. With this barrier our method becomes more similar to other predictive methods in the literature.

Even without any free parameter, our method can produce halo catalogues with the same precision of other similar methods (with free parameters) in the literature. Provided an HOD prescription, we expect to be possible to generate galaxy catalogues that have the correct linear and non-linear properties of the simulated catalogues, with $\sim 20 \%$ difference in intermediate non-linear scales.

We point out that, since the method has no free parameters and provides a clear interpretation of all steps, it can be easily extended to models beyond $\Lambda$CDM. In particular, in future works we plan to use it to create, for the first time, a method for the fast generation of halo catalogues in modified gravity, modified dark energy models and with primordial non-gaussianities. These catalogues will be key ingredients in the computation of covariance matrices in models beyond $\Lambda$CDM, allowing for precise and unbiased constraints of these models in  next-generation surveys.

\appendix
\section{Variations and code options}
\label{app:Variation}

    In this appendix we show additional results extracted from the halo catalogues that were generated with ExSHalos, using different choices of barrier and LPT orders. In Table~\ref{tab:Catalogues_test} we present all catalogues, with their specifications used in this Appendix. In all cases we have eight realizations for three different box sizes ($512$, $1024$ and $2048$~Mpc$/h$). We used the same random seeds for the four different catalogues, which allows for a one-to-one comparison.
    
\begin{table}
\centering
\caption{Specification of all catalogues used in the comparisons of Appendix~\ref{app:Variation} with their names and main characteristics.}
\begin{tabular}{l r}
\hline
Name & Characteristics \\
\hline
0LPT & Static Barrier with 0LPT \\
1LPT & Static Barrier with 1LPT \\
2LPT & Static Barrier with 2LPT \\
EB & Ellipsoidal Barrier with 2LPT \\
\hline
\label{tab:Catalogues_test}
\end{tabular}
\end{table}

\begin{figure}
\begin{center}
\includegraphics[width=\linewidth]{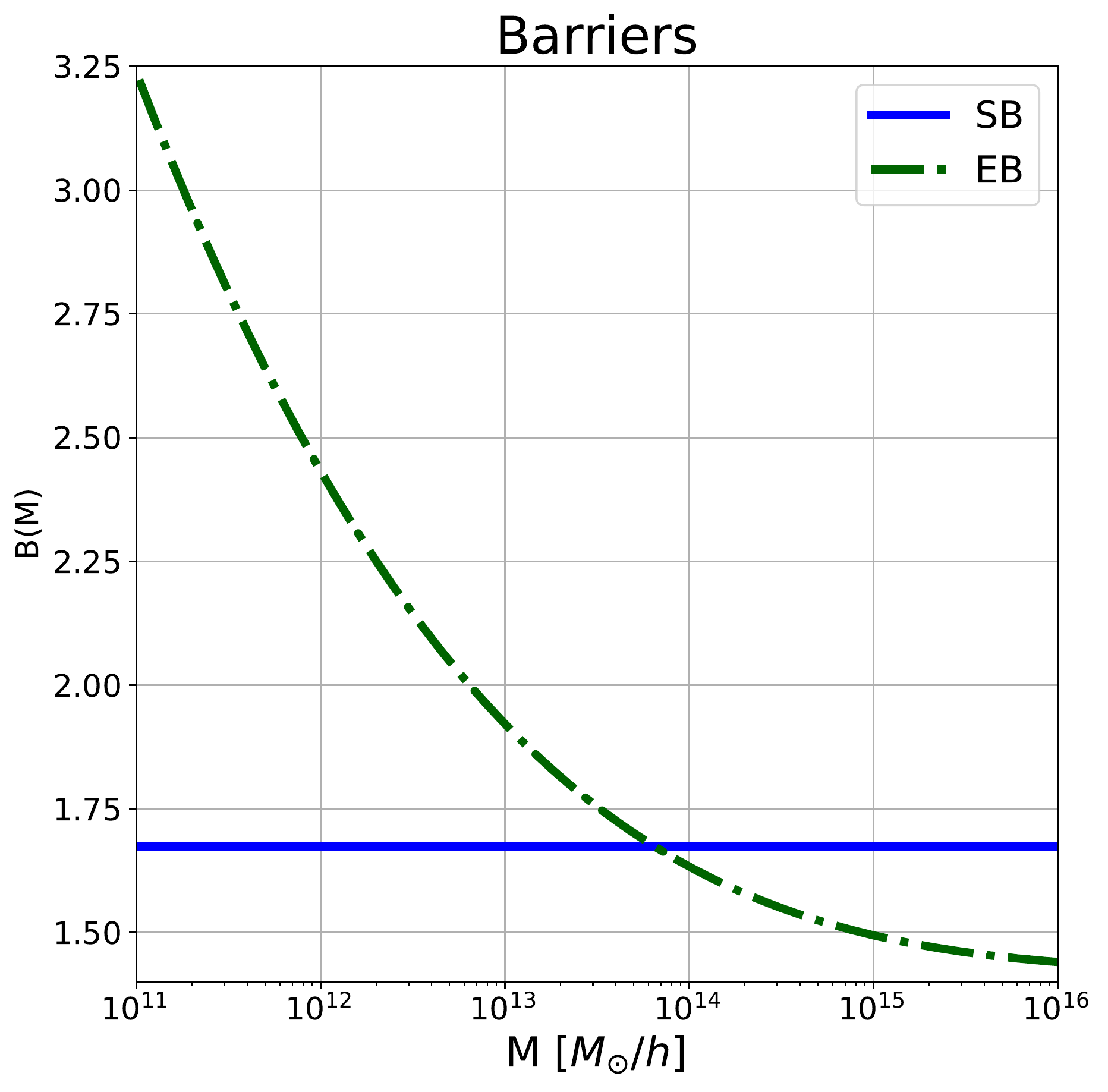}
\caption{Mass dependence of the threshold barriers $B$ used in this work to generate halo catalogues. The solid blue line shows the SB barrier for spherical collapse, and the dotted-dashed green line shows the EB barrier approximation for ellipsoidal collapse with the parameters fitted in this work.}
\label{fig:Halo_Barriers}
\end{center}
\end{figure}

    In Fig.~\ref{fig:Halo_Barriers} we show the mass dependence of the static (SB) and ellipsoidal (EB) barriers. From the shape of these barriers we can see the expected behaviour for the halo abundance, in which we get a higher number of low-mass halos when using SB, since the density threshold for their formation is lower, and conversely, we also obtain a lower number of high-mass halos, with the transition happening at masses of $\approx 7\times 10^{13} M_{\odot}/h$. These behaviours were also seen in the halo abundance of Fig.~\ref{fig:Abundance_test}.

\begin{figure*}
\begin{center}
\includegraphics[width=\textwidth]{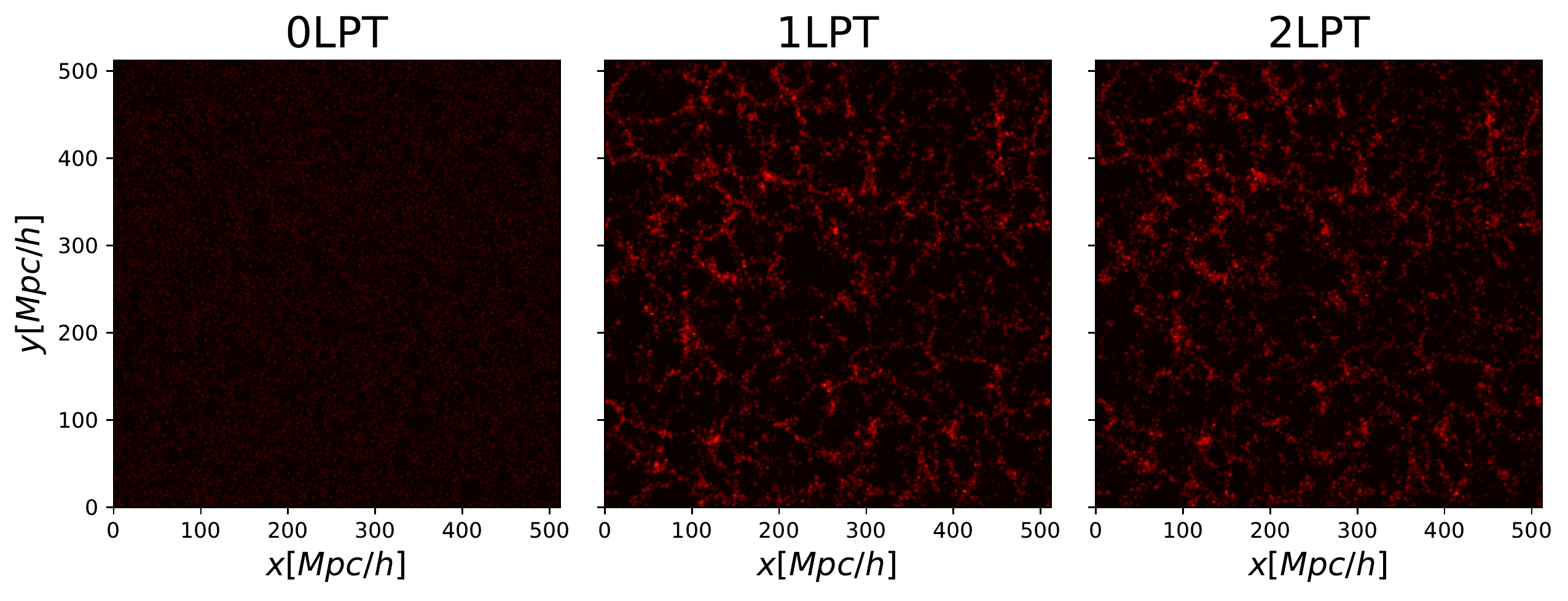}
\caption{A slice of the halo density field with width of  $8$~Mpc$/h$ in the $z$ direction, without displacing their centers (first column), displacing their centers using 1LPT (second column) and using 2LPT (third column). The same color scheme was used in the three columns.}
\label{fig:Halo_Maps}
\end{center}
\end{figure*}

    In Fig.~\ref{fig:Halo_Maps} we present the same slice of the halo density field, of width $8$ Mpc$/h$ in the $z$ direction, for three different LPT orders used in the same halo catalogue. The first column shows halos without any displacement (0LPT), the second column shows the halos displaced using LPT at first order (1LPT), and the third column shows the result of using LPT at second order (2LPT) to displace the halos. 

    The first column of Fig.~\ref{fig:Halo_Maps} shows halos found in the Gaussian map, but not displaced using LPT. As there is only one halo per cell, and they are placed at the center of their peak cell, the figure is a collection of individual points\footnote{As the slice is very thin, projection effects are not relevant.}. Differently from a particle catalogue in Lagrangian space, the halo catalogue already has some structure because not all cells have halos associated with them. This feature becomes clear when we look for the power spectra and bispectra of these catalogues, which are not constant as would be in the particle case in Lagrangian space.

    The second and third column are visually similar, which shows that second order LPT is a slight correction to first order LPT, as expected. We can also see that simply by using LPT we already have some structure in the density map, with filaments, clumps and voids in the halo map. 
    
\begin{figure*}
\begin{center}
\includegraphics[width=0.32\linewidth]{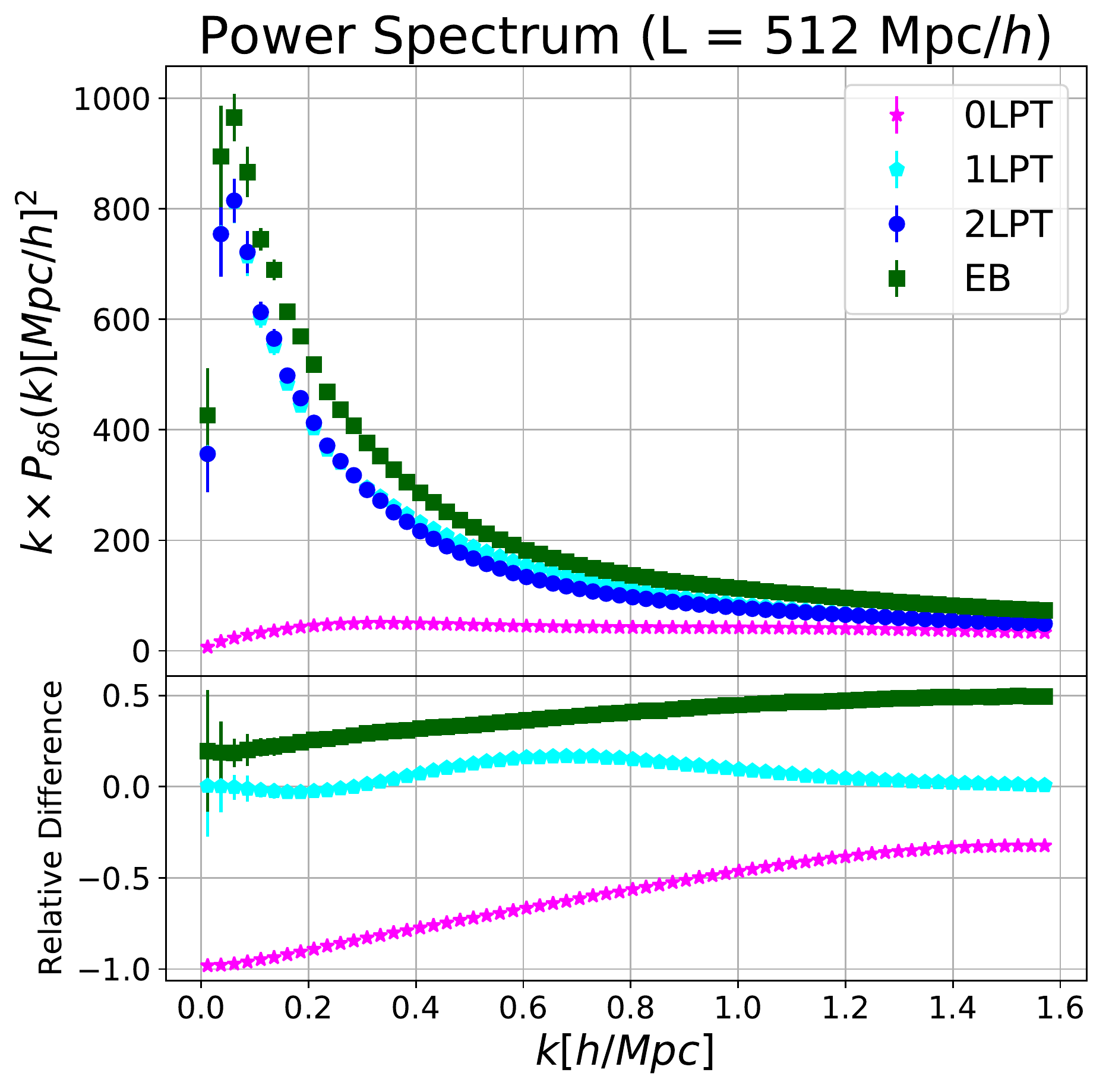}
\includegraphics[width=0.32\linewidth]{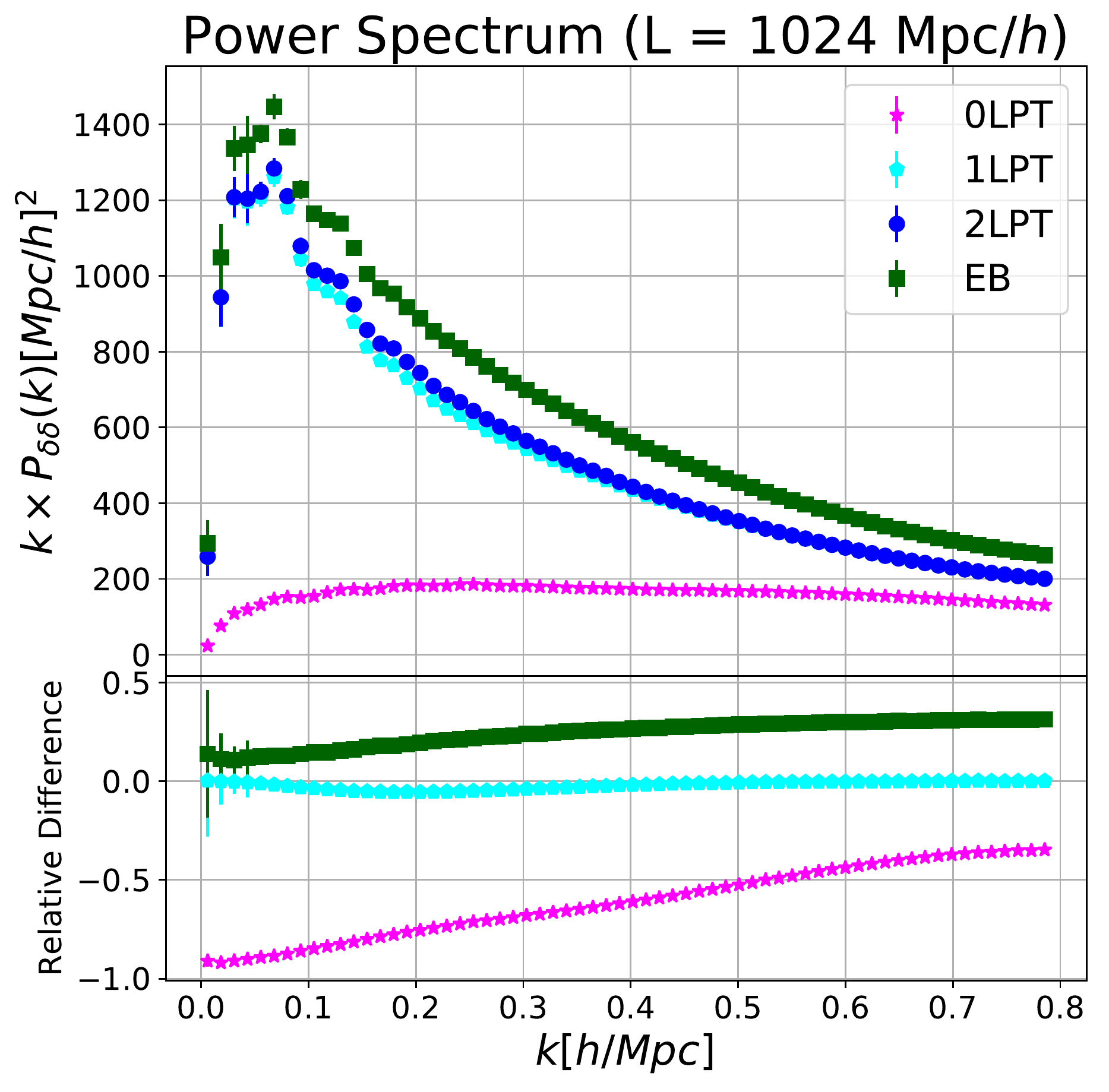}
\includegraphics[width=0.32\linewidth]{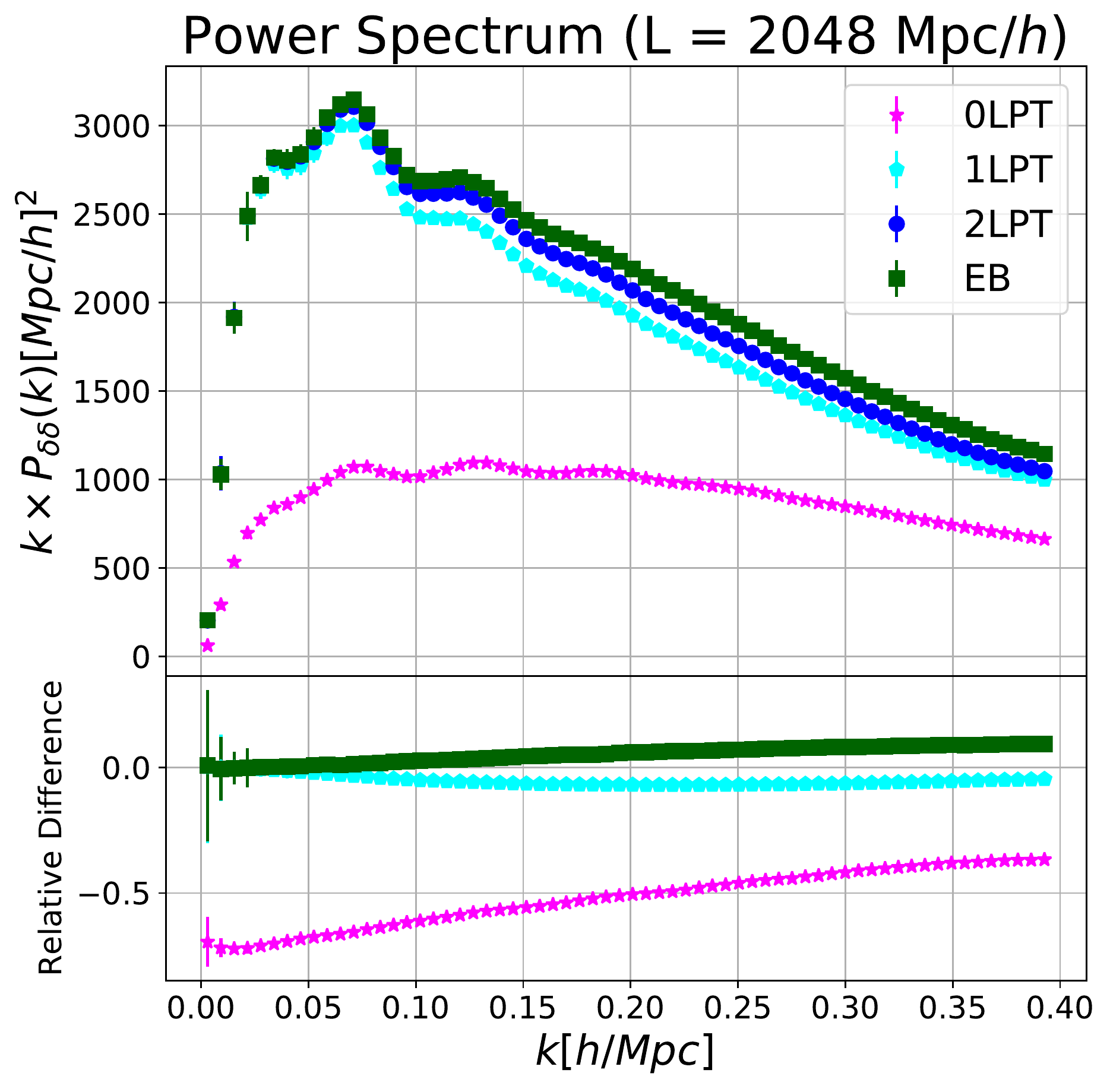} \\
\includegraphics[width=0.32\linewidth]{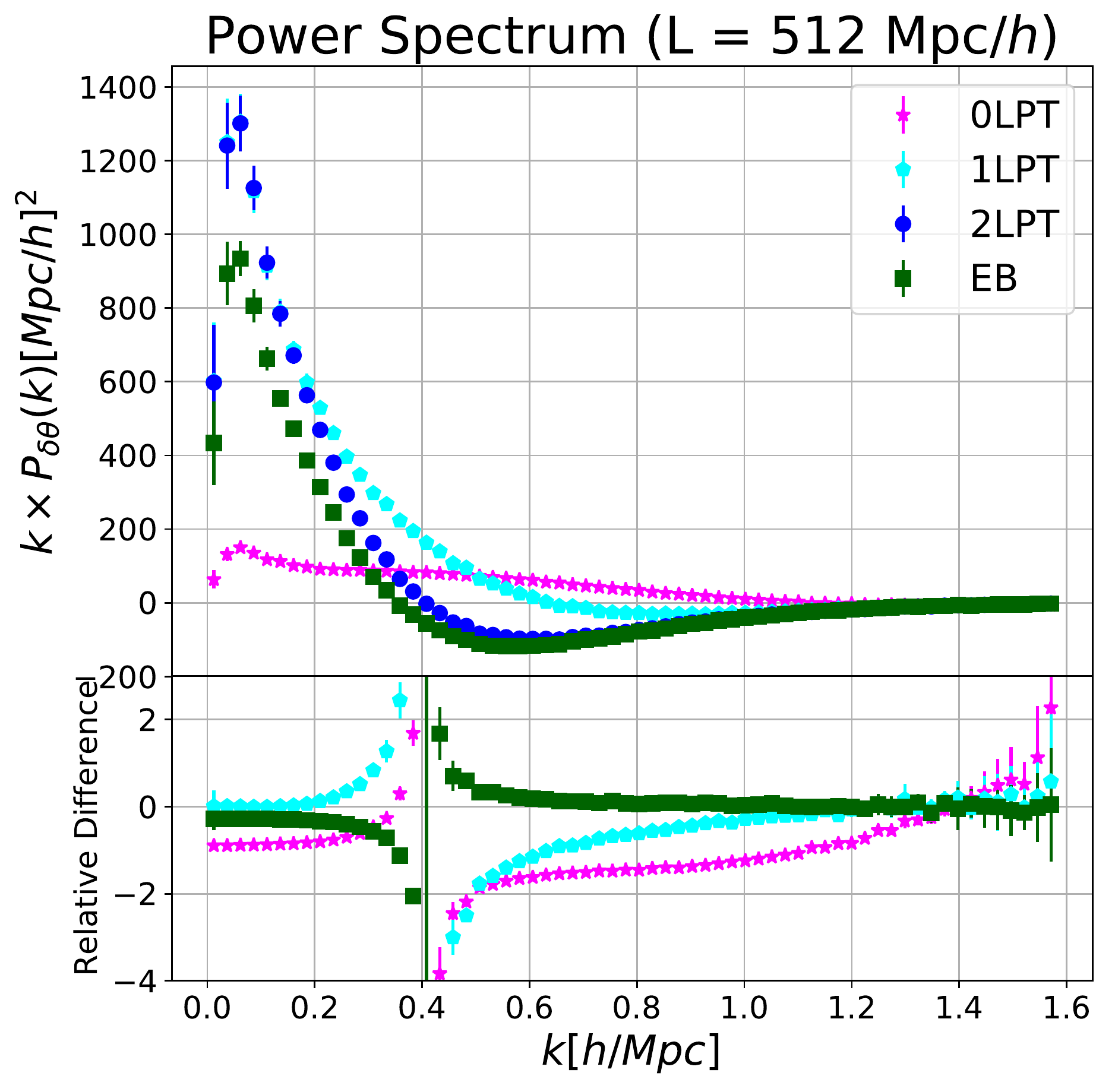}
\includegraphics[width=0.32\linewidth]{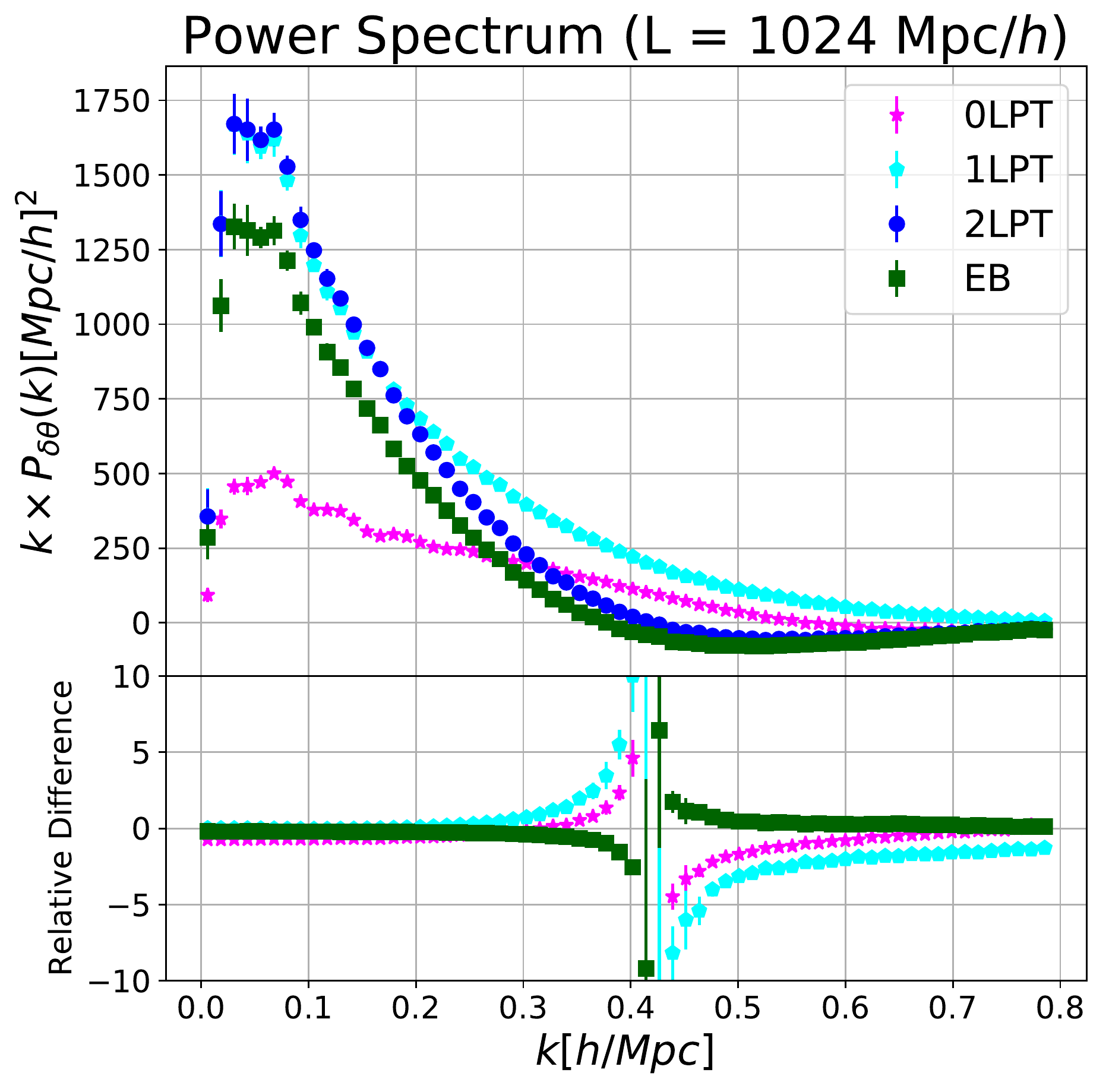}
\includegraphics[width=0.32\linewidth]{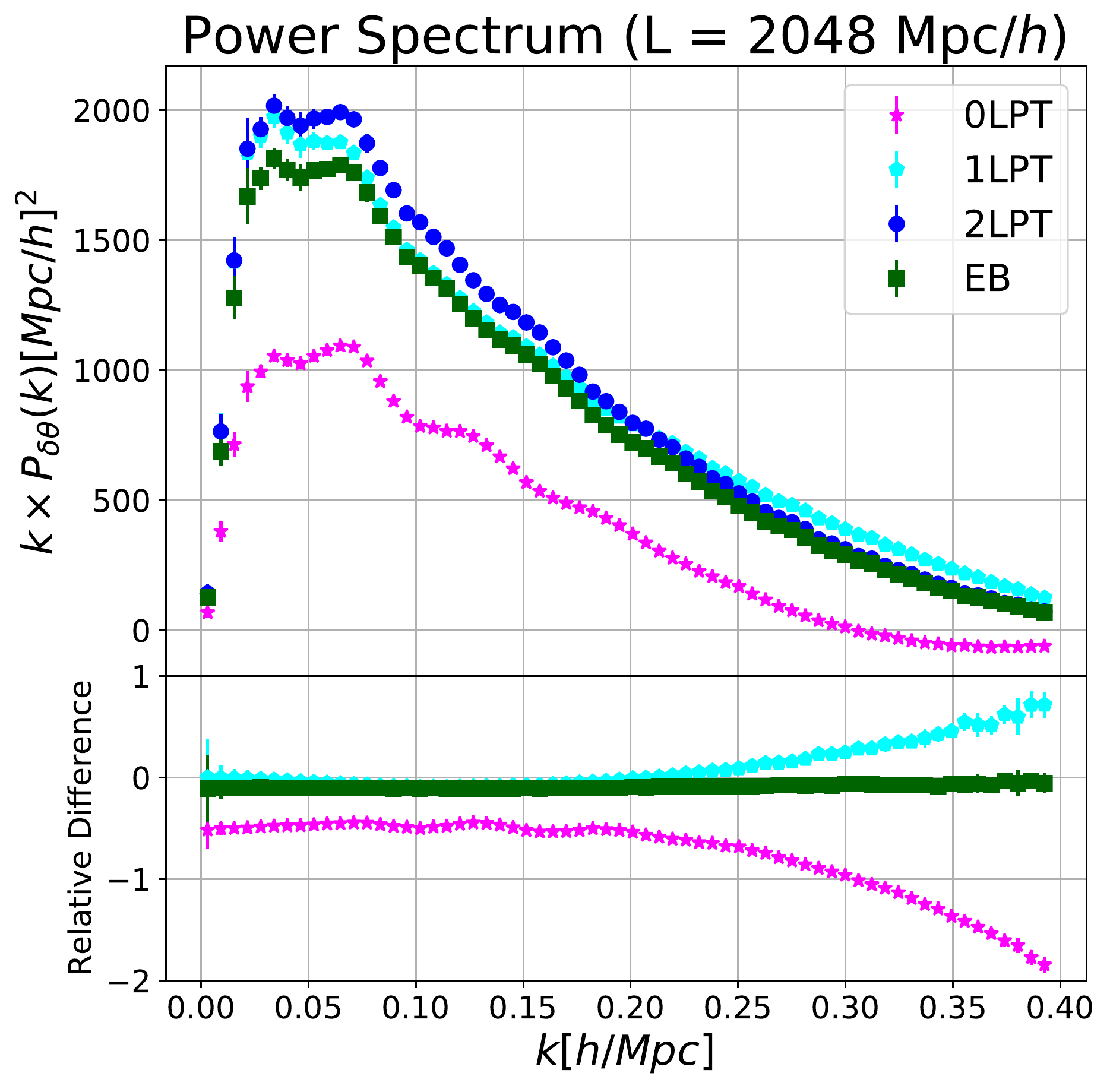} \\
\includegraphics[width=0.32\linewidth]{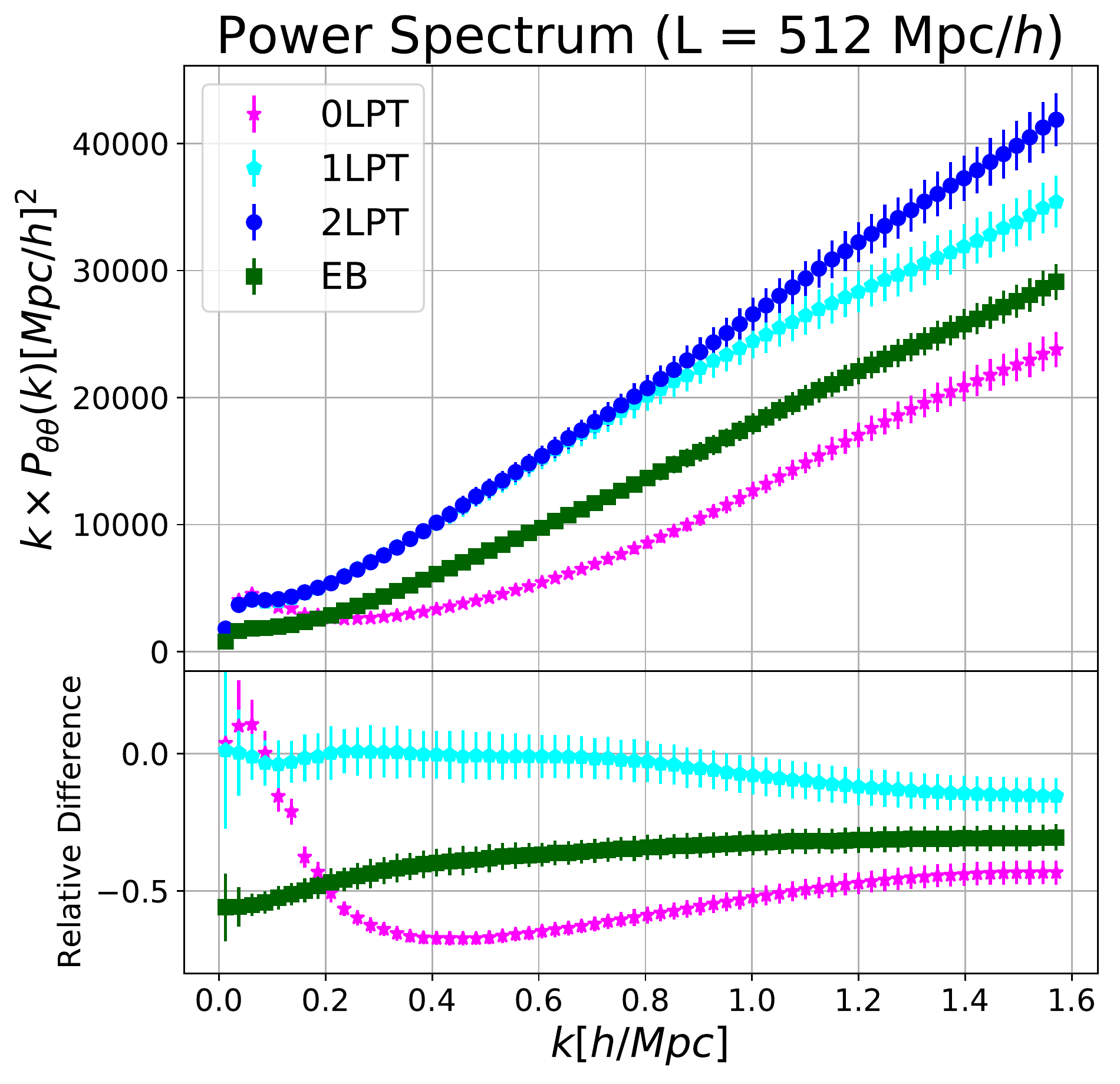}
\includegraphics[width=0.32\linewidth]{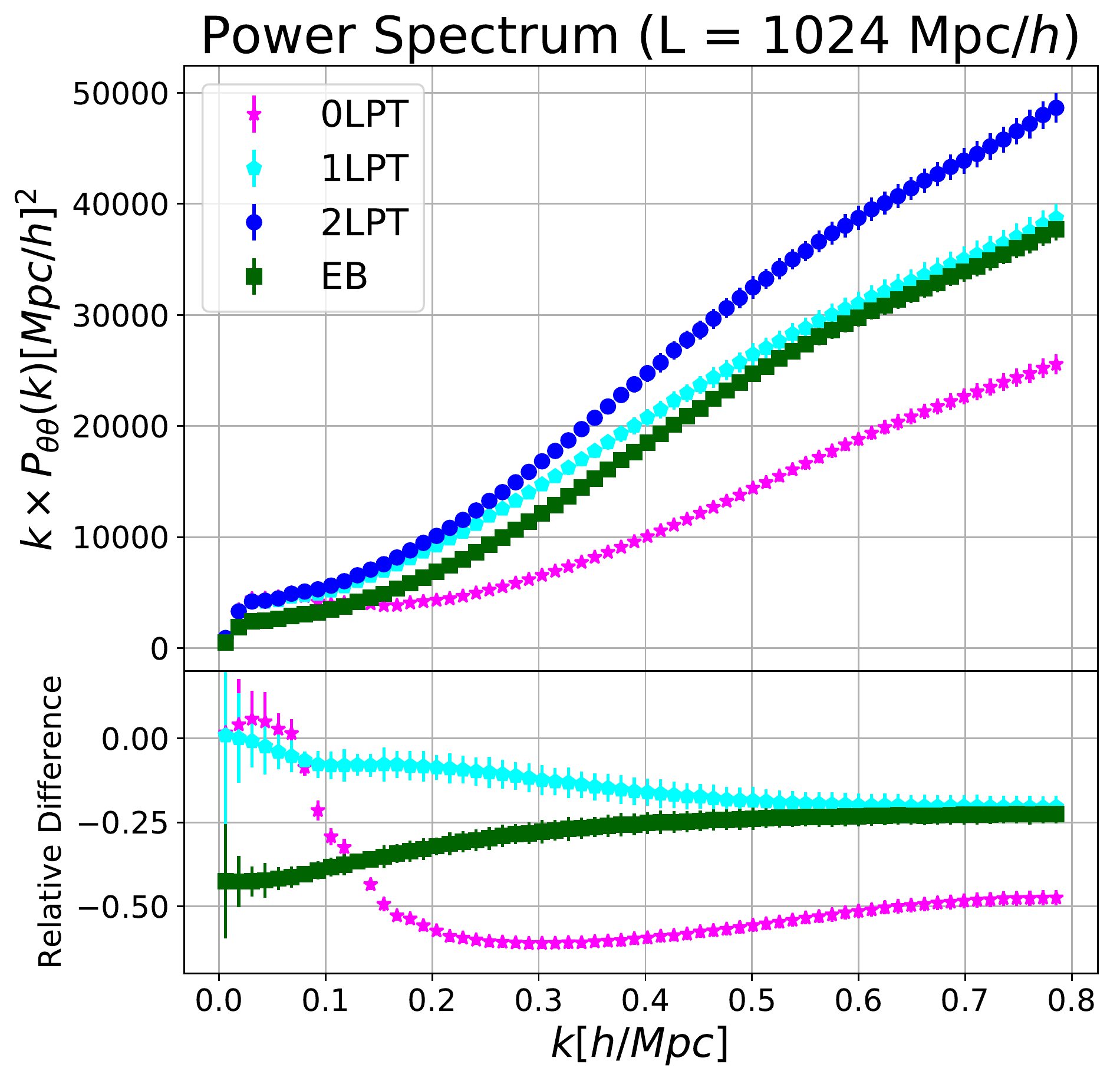}
\includegraphics[width=0.32\linewidth]{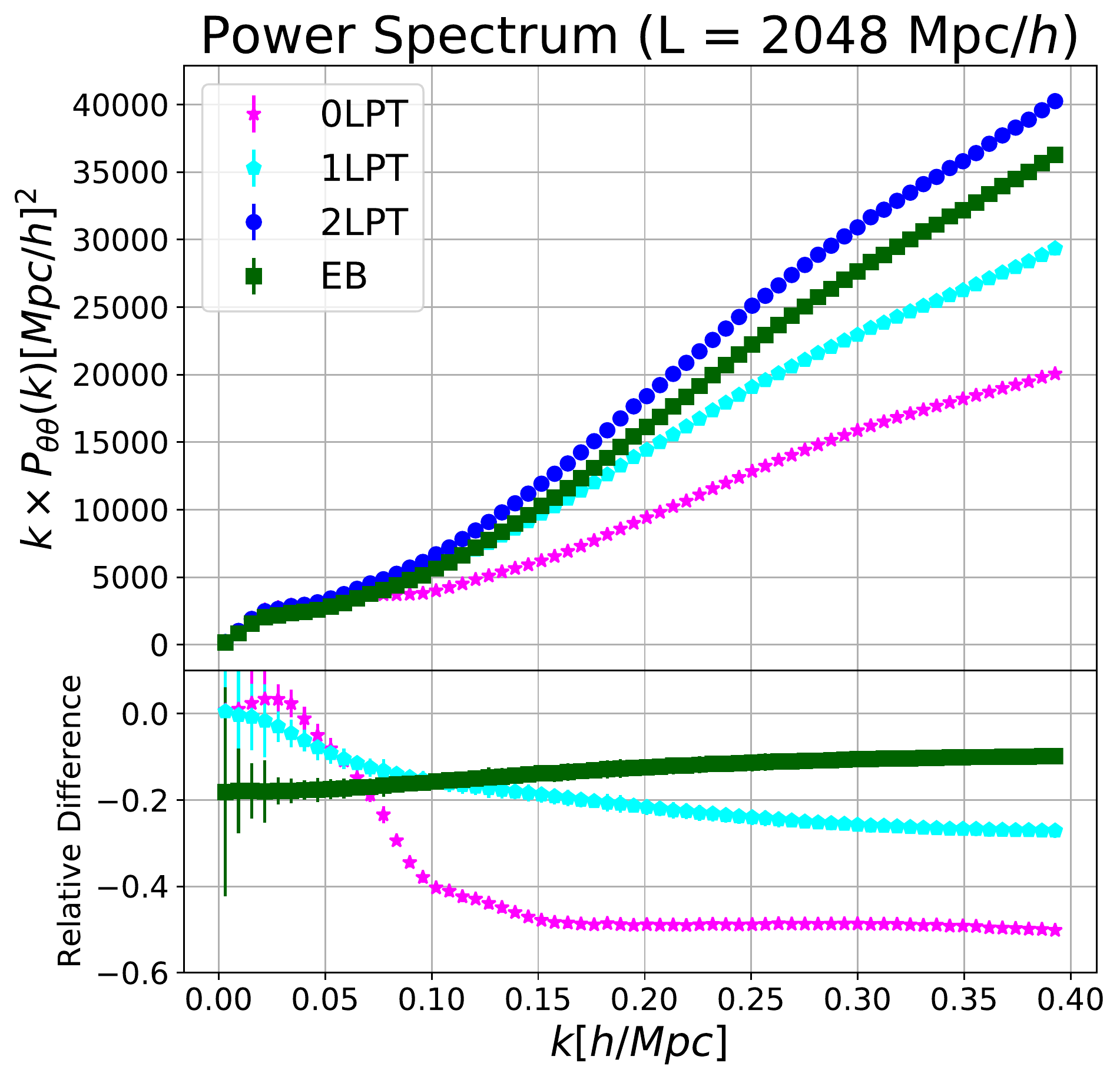}
\caption{The density-density halo power spectrum ({\it first row}), density-velocity halo power spectrum ({\it second row}) and velocity-velocity halo power spectrum ({\it third row}) for each of the three different box sizes: $L=512$ Mpc$/h$ ({\it first column}), $1024$ Mpc$/h$ ({\it second column}) and $2048$ Mpc$/h$ ({\it third column}). We present the mean of each observable for the eight realizations, with $\pm 1 \sigma$ errors. Results are shown for the SB halo catalogues with 2LPT (blue dots), 1LPT (cyan pentagons) and 0LPT (pink stars), along with the EB catalogues with 2LPT (green squares). Bottom panels present the relative differences between each catalogue and the SB catalogue with 2LPT.}
\label{fig:Powers_Barriers}
\end{center}
\end{figure*}

\begin{figure*}
\begin{center}
\includegraphics[width=0.32\linewidth]{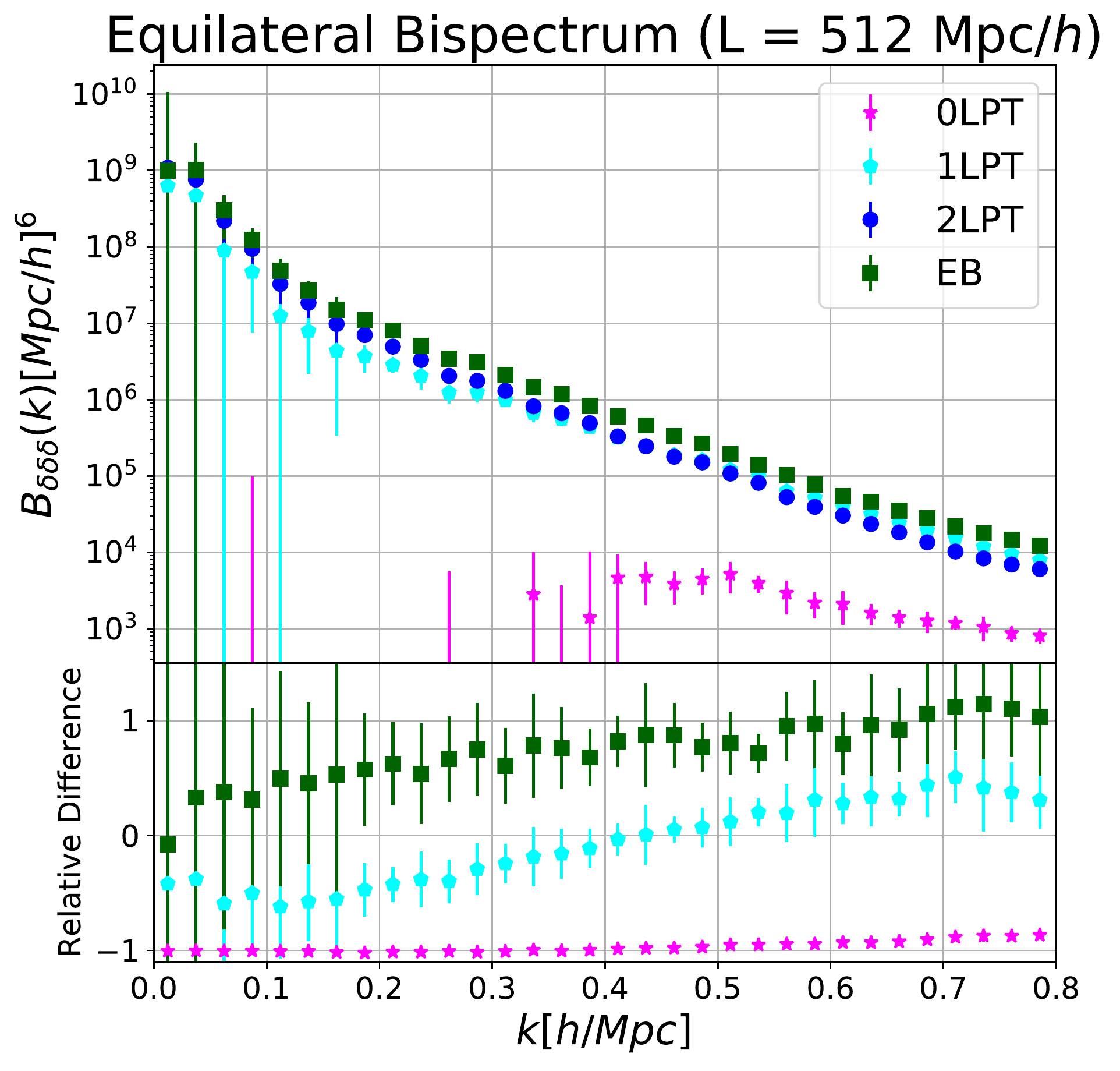}
\includegraphics[width=0.32\linewidth]{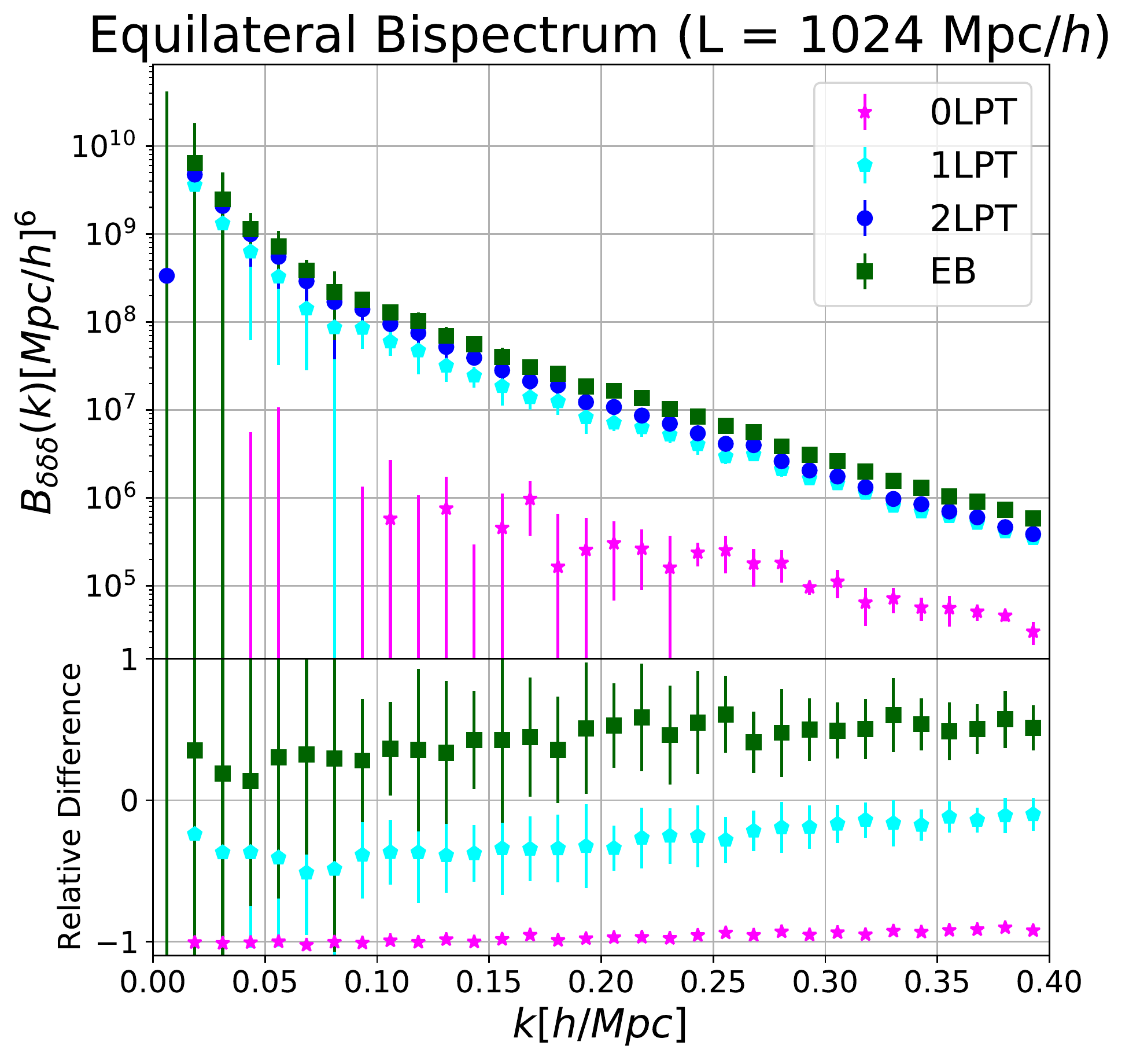}
\includegraphics[width=0.32\linewidth]{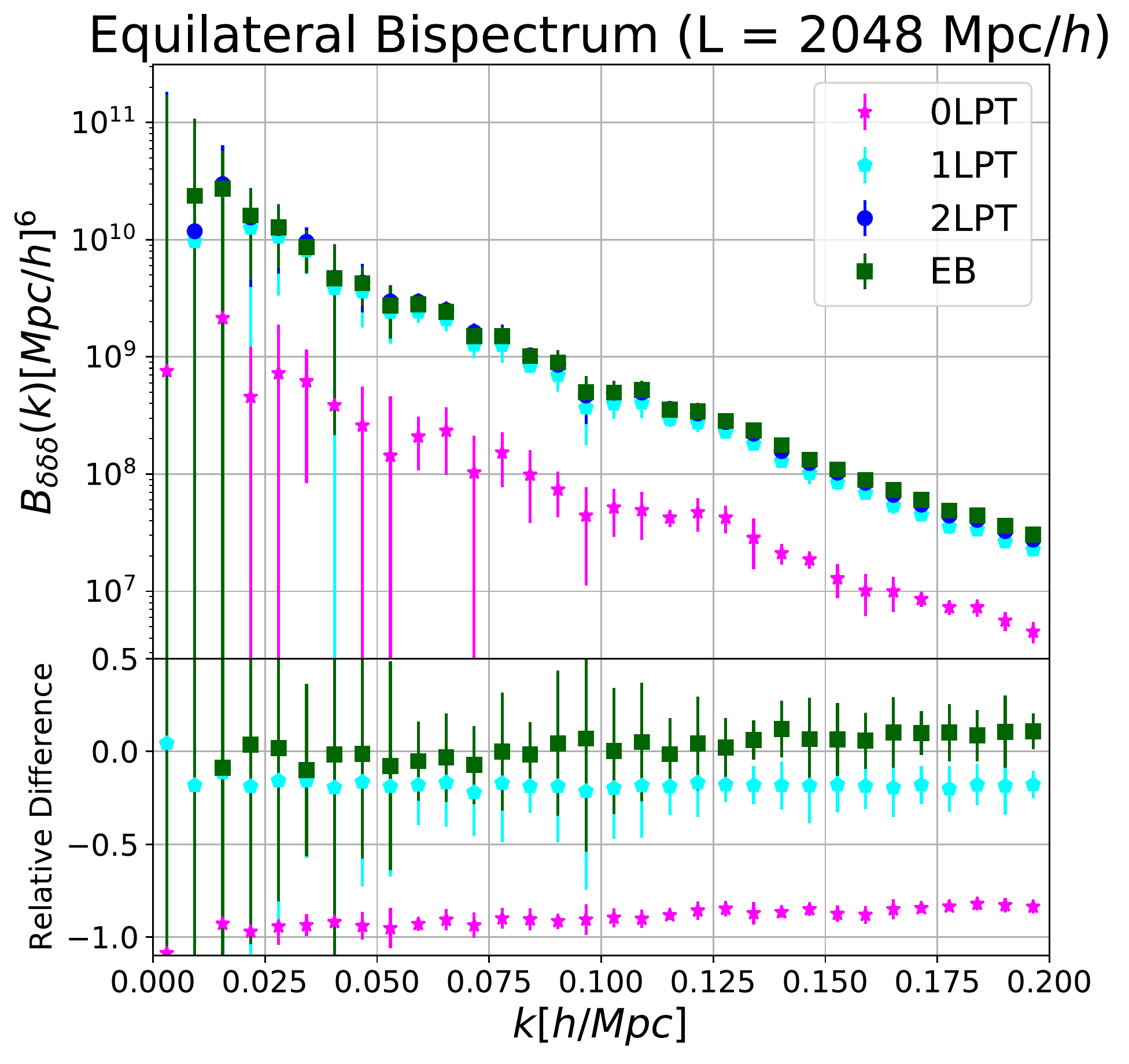} \\
\includegraphics[width=0.32\linewidth]{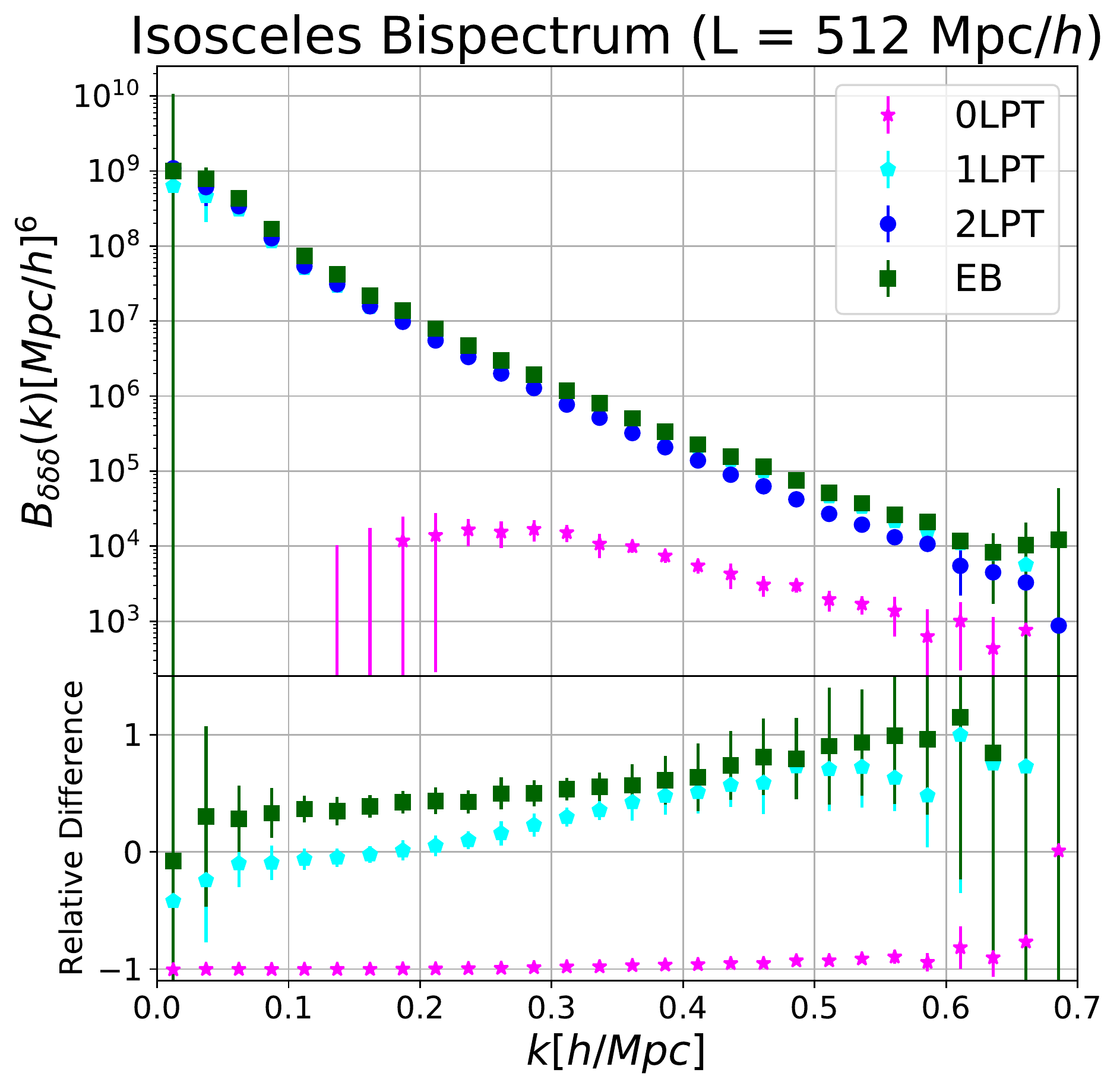}
\includegraphics[width=0.32\linewidth]{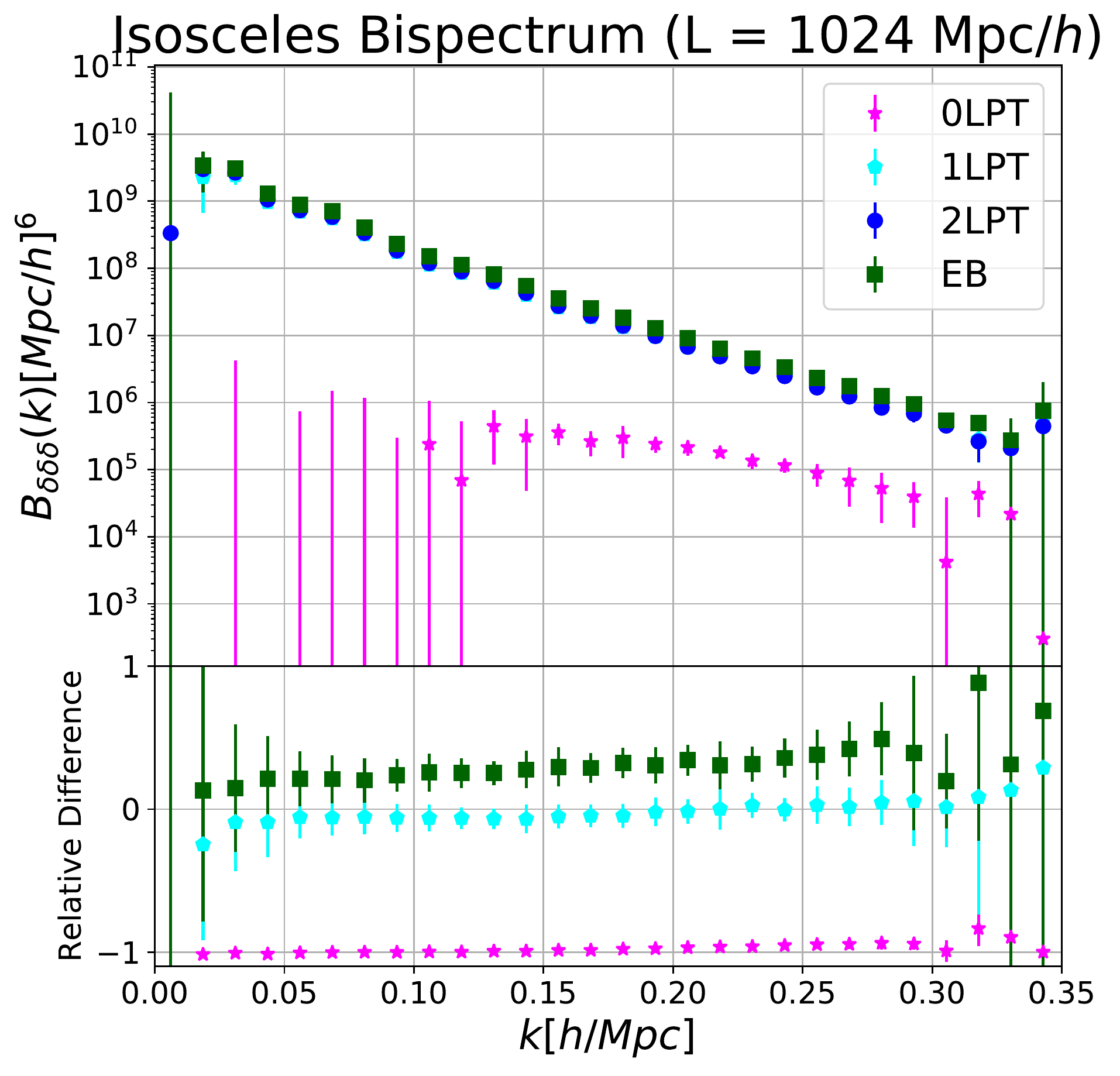}
\includegraphics[width=0.32\linewidth]{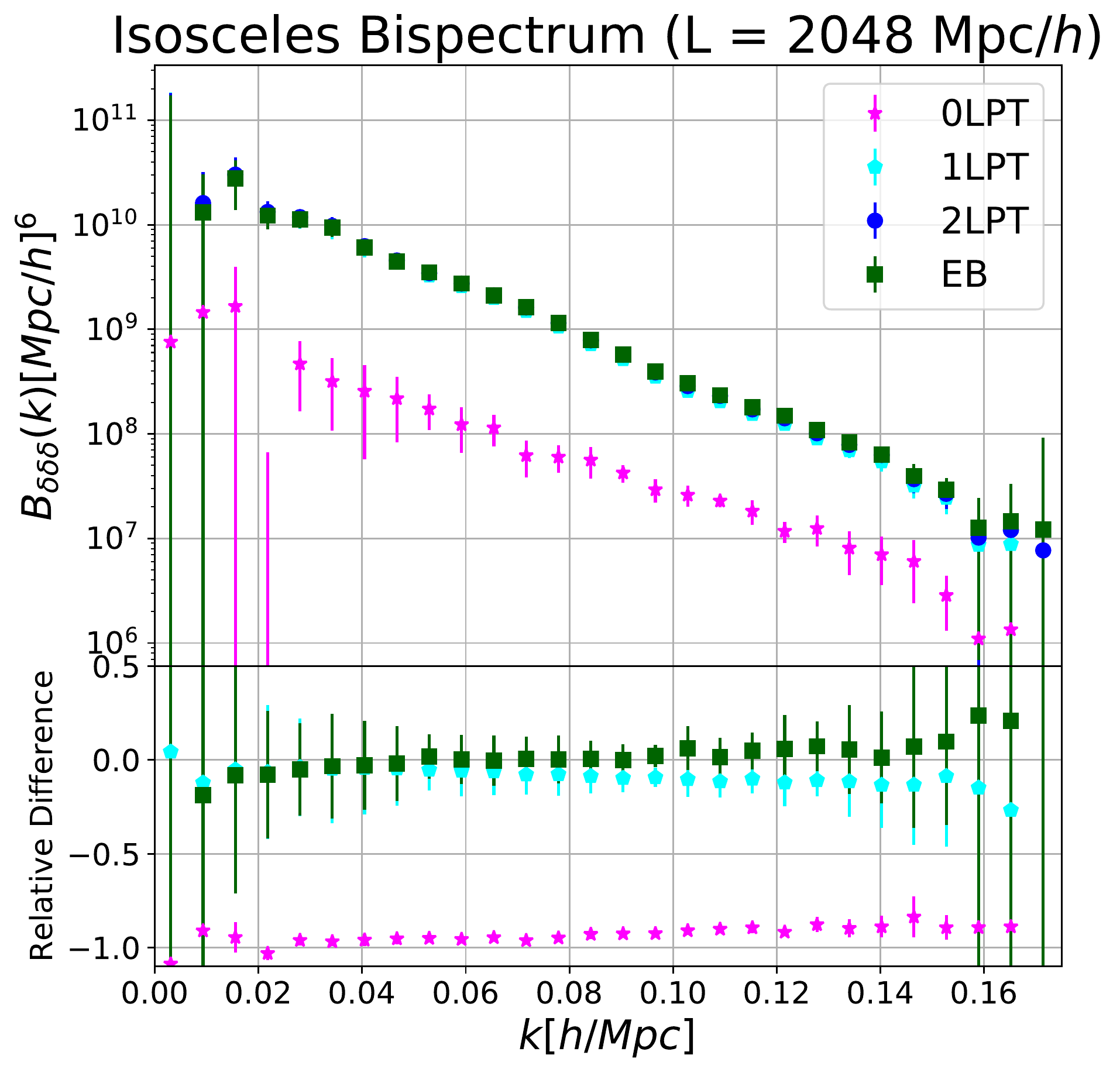} \\
\includegraphics[width=0.32\linewidth]{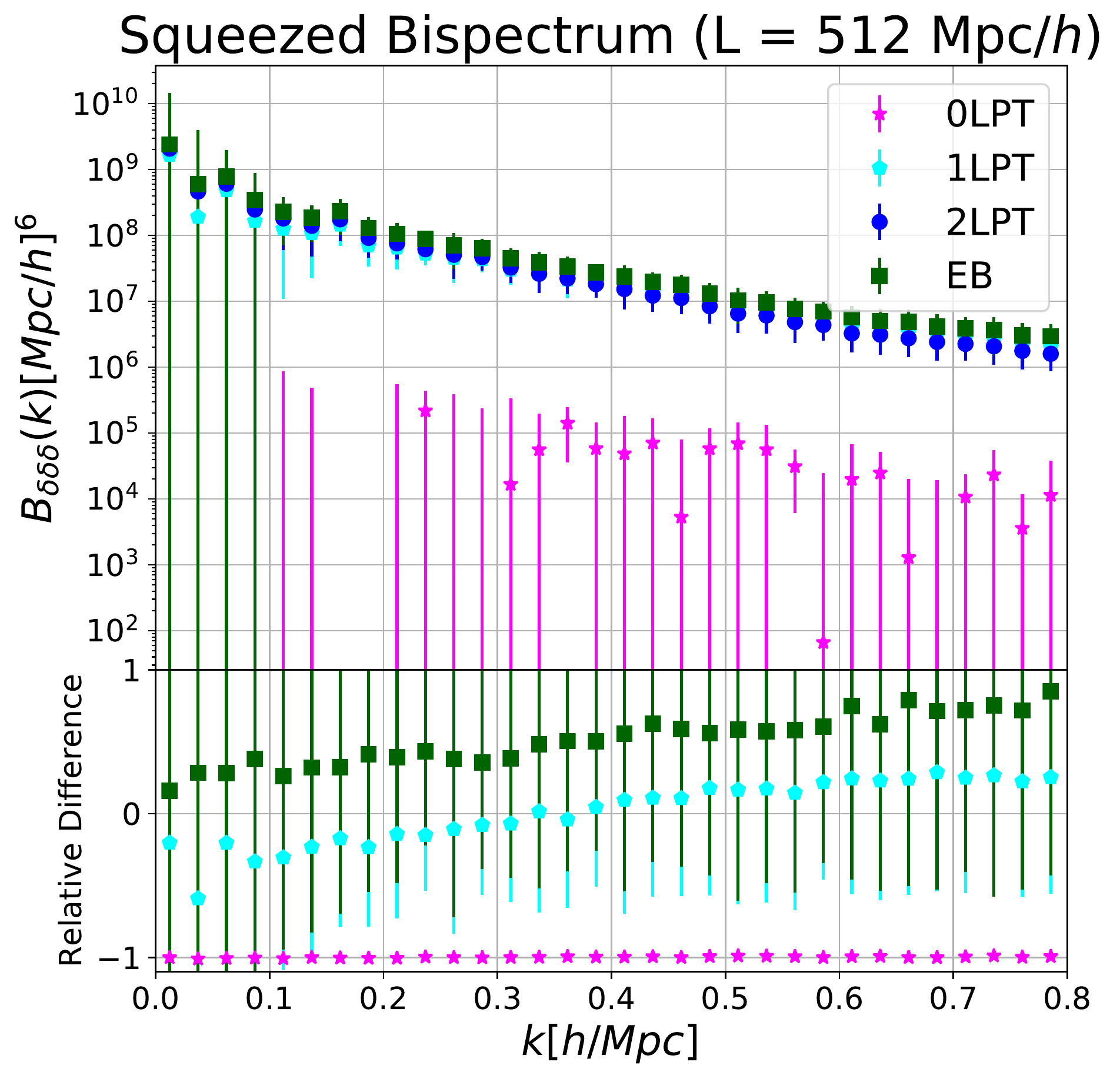}
\includegraphics[width=0.32\linewidth]{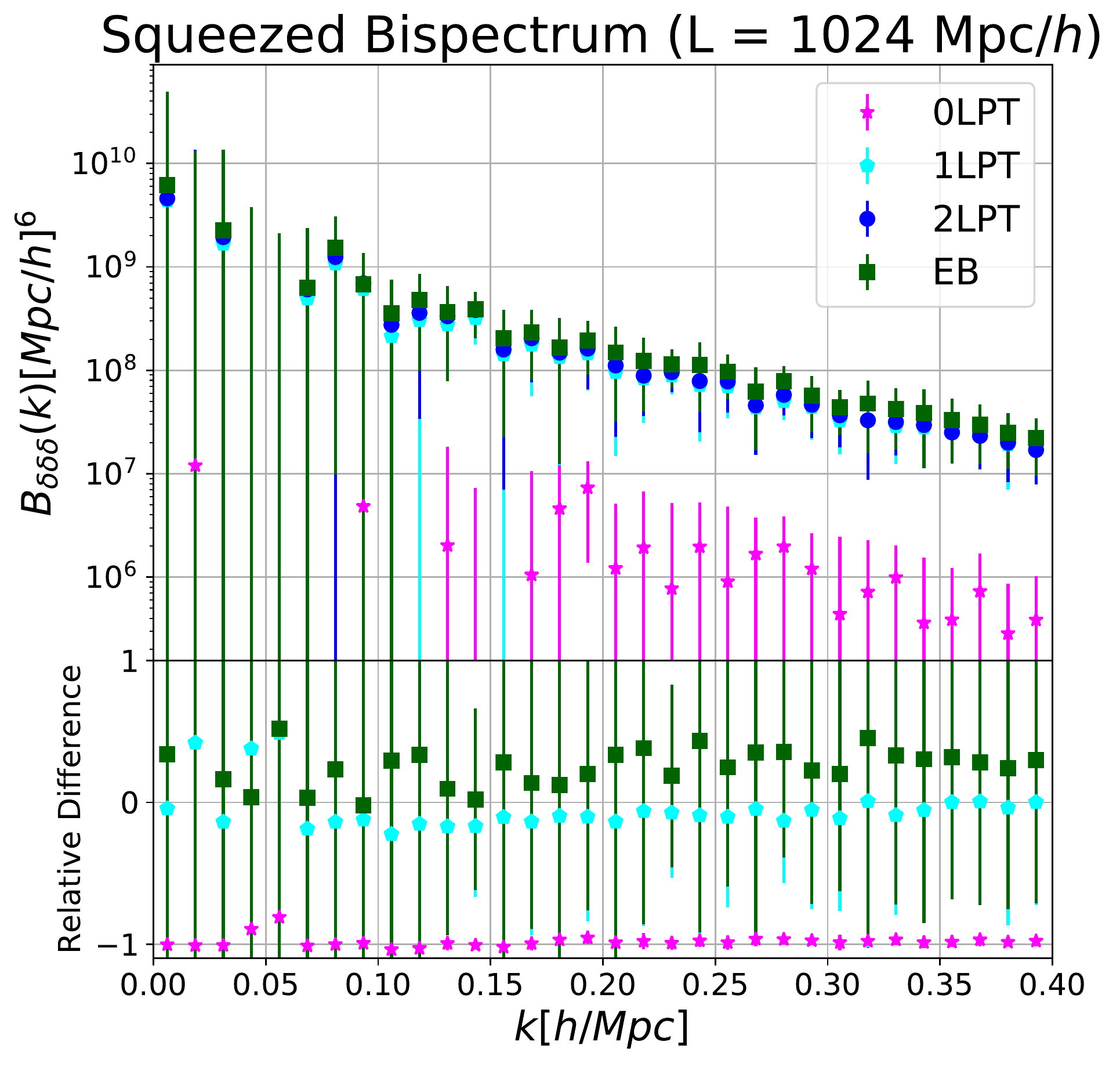}
\includegraphics[width=0.32\linewidth]{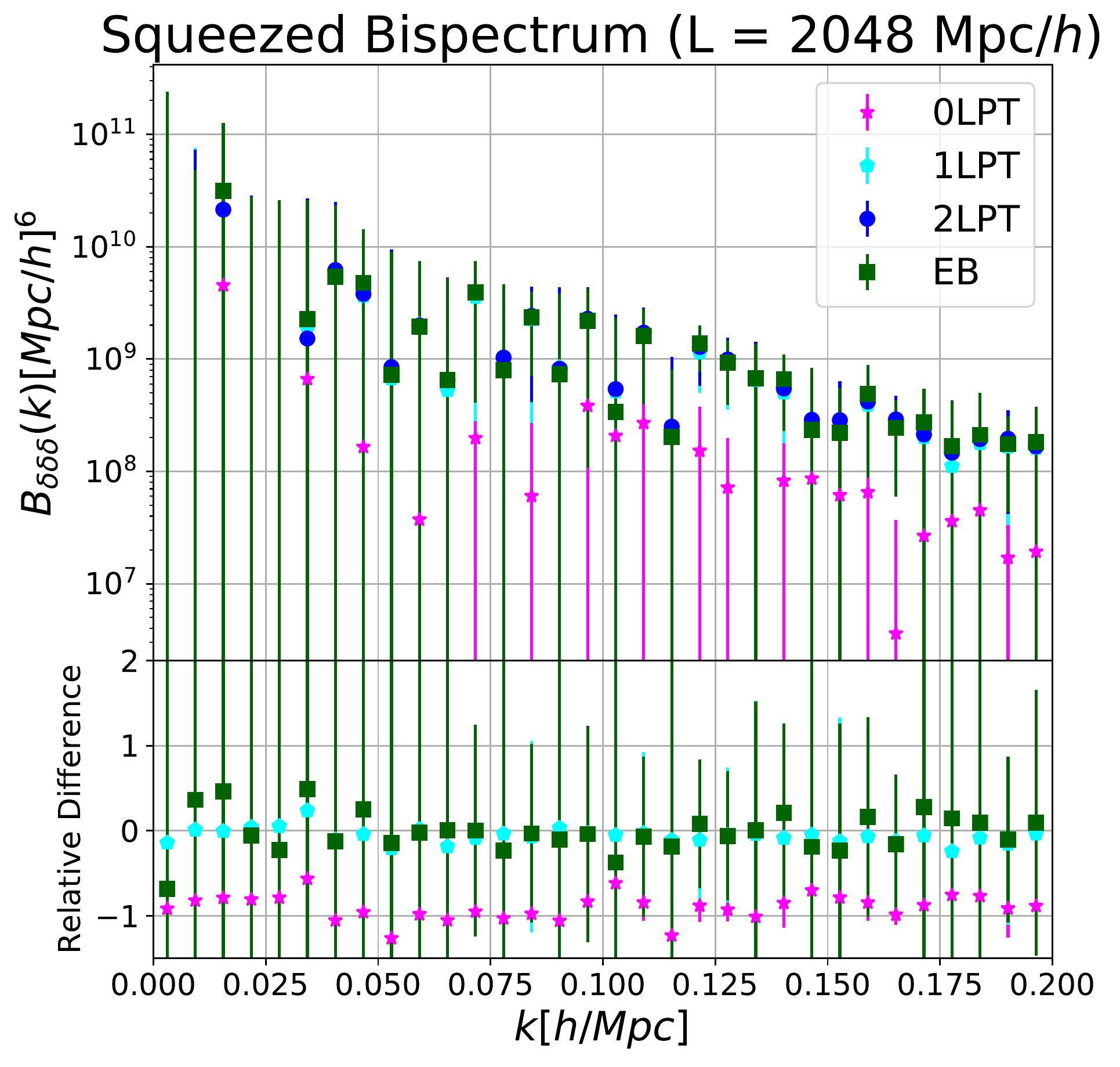}
\caption{The density-density-density halo bispectrum for the equilateral triangles ({\it first row}), isosceles triangles ({\it second row}) and triangles in the squeezes limit ({\it third row}) for each of the three different boxes size: $L=512$ Mpc$/h$ ({\it first column}), $1024$ Mpc$/h$ ({\it second column}) and $2048$ Mpc$/h$ ({\it third column}). We present the mean of each observable for the eight realizations, with $\pm 1 \sigma$ errors. Results are shown for the SB halo catalogues with 2LPT (blue dots), 1LPT (cyan pentagons) and 0LPT (pink stars), along with the EB catalogues with 2LPT (green squares). Bottom panels present the relative differences between each catalogue and the SB catalogue with 2LPT.}
\label{fig:Bis_Barries}
\end{center}
\end{figure*}

    In Figs.~\ref{fig:Powers_Barriers} and \ref{fig:Bis_Barries} we show a quantitative comparison between the four catalogues presented in Table~\ref{tab:Catalogues_test}. The density-density, density-velocity and velocity-velocity power spectra are shown in Fig.~\ref{fig:Powers_Barriers}, while the bispectra for our three triangular configurations are shown in Fig.~\ref{fig:Bis_Barries}. We also show separately the results for each box size in each column. We used $256$ cells per dimension to measured the power spectra and $128$ cells to measured the bispectra, and consider all halos with more than $8$ particles (this is because we are using realizations with the same initial conditions and the halo completeness is not relevant).
    
    In the second and third rows of Fig.~\ref{fig:Powers_Barriers}, we assigned the velocity given by 1LPT for the 0LPT catalogues, so that the halos with no position displacement could still have some non-zero velocity associated to them.  
    The pink stars in these plots show the power spectra between the 0LPT density map and the 1LPT velocity map. The 0LPT and 1LPT results, in the third row of this figure, are different because the halos are in different places in both maps, and this changes the velocity power spectrum even with halos being the same, and having the same velocities.
    
    In these figures it is possible to observe the effect of LPT order when we displace halo centers. In particular, comparing the results of the 0LPT catalogue with the results of the 1LPT and 2LPT catalogues, we can see the importance of these displacements: when we do not displace the halos we have much lower amplitudes in the power spectra, as well as bispectra which nearly vanish. The effect of LPT is more relevant for small halos, because the displacements are much larger than the halo radii -- we can see this in the first column of both figures.
    
    Another interesting feature that happens for halos, but not for particles, is that even 0LPT halos have non-vanishing power spectra and bispectra, despite being much smaller than for the 1LPT and 2LPT cases. The non-vanishing density power spectrum also generates the non-vanishing linear halo bias observed in Fig.~\ref{fig:Bias_test}. This shows that the halo distribution already presents some structure even in Lagrangian space, as expected by the excursion set theory (see \cite{deSimone2} for a review of the excursion set calculation for the halo bias).
    
    Considering now the differences between the 1LPT and 2LPT catalogues, we can see the effects of the second order contributions in the halo catalogues\footnote{Remember that we are using the same initial conditions for the four ExSHalos catalogues.}. First, we see that at linear scales the two catalogues agree, which is expected by the LPT. The differences in the density power spectra are also very similar for the three box sizes, with only minor differences in the small and large boxes. On the other hand, the velocity-velocity power spectra present large differences at non-linear scales, with less power in 1LPT catalogues, in a manner  similar to the comparison with the simulations and the SB catalogues in Fig.~\ref{fig:Ptthalos_NBody}. This shows that non-linearities are more pronounced in the velocity field than in the density field. 
    Non-linearities are also more relevant for the most massive halos, which are in the densest regions and generate the highest velocity fields.
    
    When we compare the power spectra (in the first row of Fig.~\ref{fig:Powers_Barriers}) with the bispectra (in Fig.~\ref{fig:Bis_Barries}) it is possible to see that the second-order term in LPT is more important for 3-point correlation functions. This happens because the higher-order correlation functions are better suited to capture information about non-linear structures such as filaments, sheets, halos and voids.
    
    In both figures we can see the effect of halo definitions in the Fourier-space correlation functions. The first difference is the linear behaviour of the density power spectra (first row of Fig.~\ref{fig:Powers_Barriers}), the linear halo bias. This difference is expected from theory when we use two different barriers to compute the linear halo bias.
    
    The $k$-dependence of power spectra and bispectra is very similar for the two catalogues, as noted for the two halo definitions in the simulations (see for instance Fig.~\ref{fig:Phalos_NBody}). This shows that the barrier parameters affect only slightly the shape of the correlation functions, affecting mainly the overall normalization. This happens because the barrier choice changes the conditional probability for halo formation, which changes all bias coefficients, not only the linear one. On the other hand, the linear coefficient is much more important than the higher order terms, which could change the shape of the spectra (see \cite{deSimone2} for a discussion about the computation of non-linear coefficients and \citep{Abidi} for a discussion of the dependence of the halo power spectrum on these coefficients).
    
    In summary, we clearly need to displace halos using LPT in order to recover the correct halo power spectra and bispectra, but we need the second order corrections only if we are interested in higher precision in 3-point correlation functions, as well as higher order correlation functions. The choice of the barrier parameters change the correlation functions mainly on linear scales, and affect only slightly their $k$-dependence.

\section{One-to-one Comparison}
\label{app:One}

    In this appendix we consider a more direct one-to-one comparison between our method and a N-body simulation. For this comparison, we ran the \texttt{RAMSES} code \citep{Teyssier} as well as our method using the same initial conditions. In both cases, we used a box with size $128$ Mpc$/h$ with $256$ cells per dimension. We generated both outputs at $z=0$, and found the halos in the simulation using a spherical overdensity code with $\Delta = 360$. For the halo catalogue generated with our method, we used the SB case with 2LPT displacement. 
    
    Generating the halo catalogue for the N-body simulation took us approximately one day using $24$ processors, while for our method it took approximately $30$ seconds using only $1$ processor. Therefore, ExSHalos is $\sim 35,000$ times faster than the simulation\footnote{We consider that the $24$ processors improve the computational time by a factor of $12$ for this simulation, by taking into account the time lost in memory sharing in a conservative way.}.
    
\begin{figure}
\begin{center}
\includegraphics[width=\linewidth]{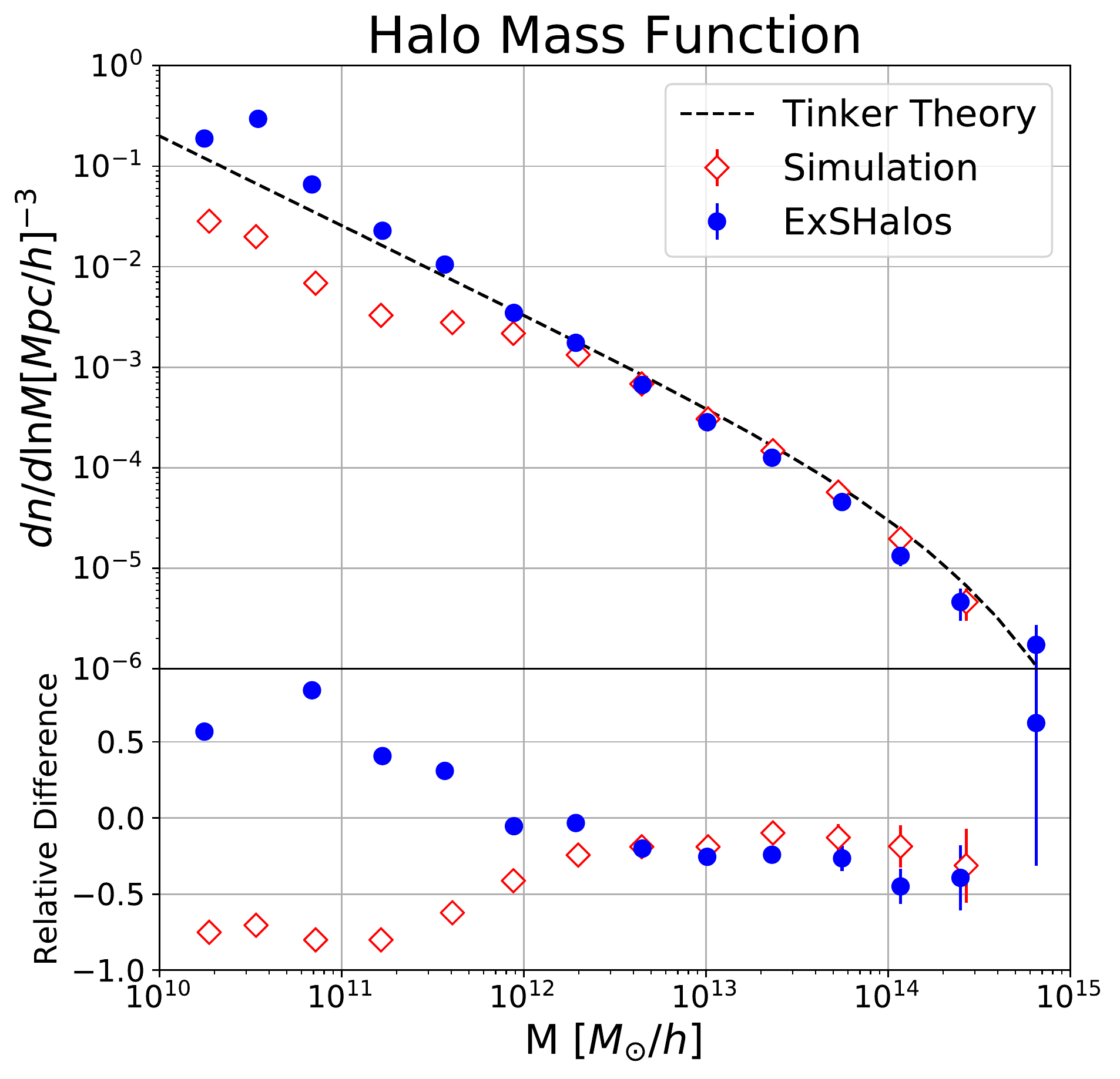}
\caption{Halo abundance for the simulated catalogue (red diamonds) and for the catalogues generated with our method using the SB model and 2LPT displacement (blue dots), both using the same initial conditions. We also show the fitting formula from \citep{Tinker} (dashed line) for comparison. The bottom panel shows the relative difference of each catalogue with respect to Tinker's prediction.}
\label{fig:Halo_Abundance_one}
\end{center}
\end{figure}

    In Fig.~\ref{fig:Halo_Abundance_one} we show the abundance of halos found in the N-body simulation and in the halos generated using our method with the same initial conditions. Here we show all halos found in these catalogues without imposing any mass cut, which therefore displays the low-mass resolution properties of each catalogue. We also show the fitting formula from \cite{Tinker} for reference.
    
	Fig.~\ref{fig:Halo_Abundance_one} shows the same feature seen in Fig.~\ref{fig:Abundance_NBody}, where the halos generated with our method have a larger abundance at small masses and a lower abundance at large masses. At low masses, the lack of abundance for the simulated halos is in contrast with an excess for ExSHalos. In the simulations this happens because we do not have enough resolution to map the density field of low-mass halos. In ExSHalos this happens because we attribute all density peaks to small halos, even when they are not real halos. 
	
	Therefore, in comparison with N-body simulation halos, our method gives a better description of the density peaks up to the mass resolution of the underlying matter density field. As for the generation of galaxy catalogues, we are mainly interested in the density peaks, so our method could be used with a resolution lower than full N-body simulated halo catalogues, implying even less computational cost.

\begin{figure*}
\begin{center}
\includegraphics[width=0.8\textwidth]{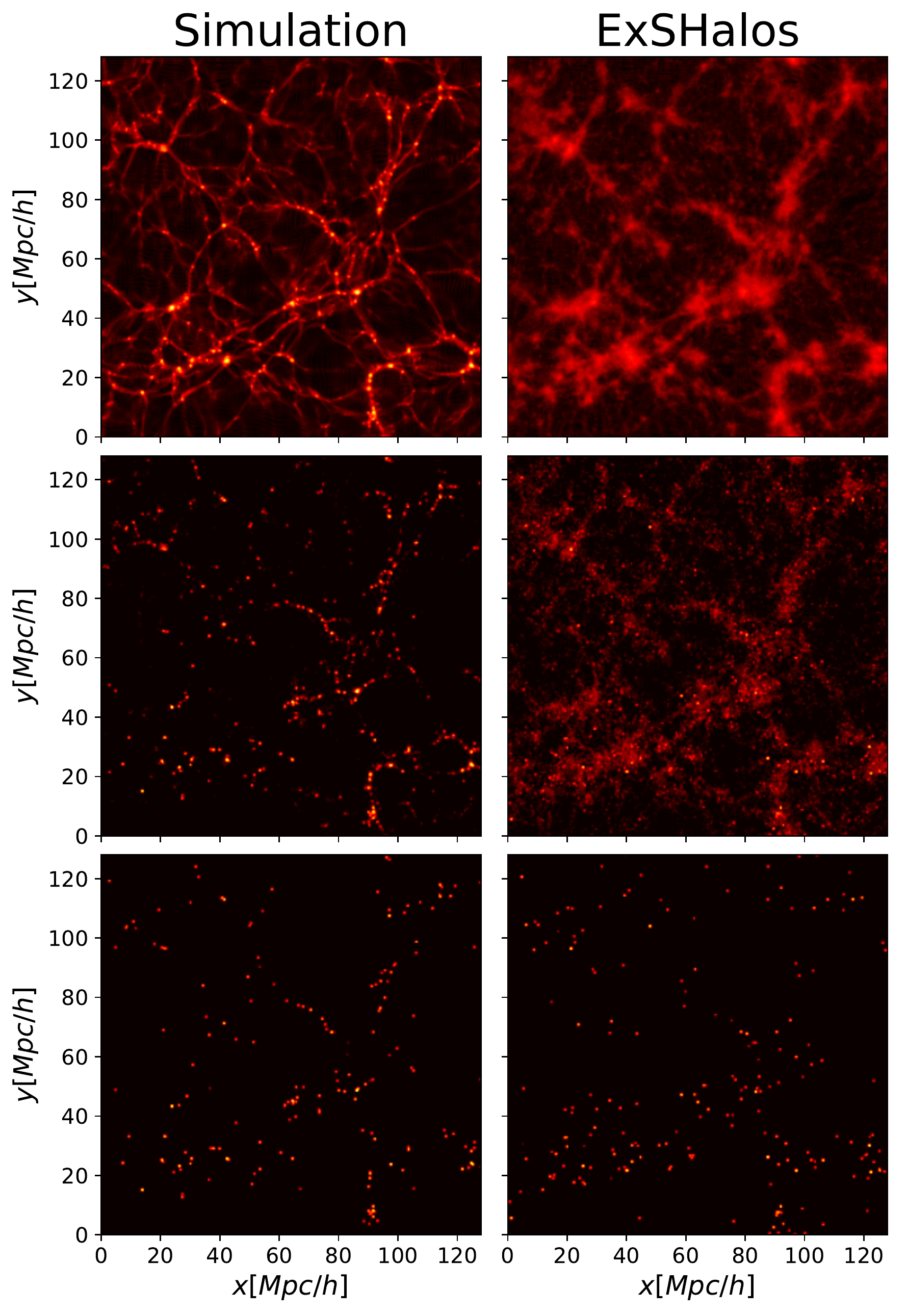}
\caption{Comparison between the simulation ({\it left column}) and our method with the SB model and 2LPT ({\it right column}), both using the same initial conditions. The panels show density maps for a slice of $4$ Mpc$/h$ in the $z$ direction. The first row show the final density map ($z=0$) for the particles in the simulation and for our 2LPT. The second row shows the density map for all halos found in both cases, and the third row shows the density map for all halos with more than $100$ particles. Dark areas represent underdensities (voids) and red areas map the densest regions.}
\label{fig:Halo_Maps_one}
\end{center}
\end{figure*}

    Fig.~\ref{fig:Halo_Maps_one} displays a qualitative comparison between some steps of both codes, showing the density maps for a slice of $4$~Mpc$/h$ in the $z$-direction. The first row shows the final matter catalogue for the simulation along with the density catalogue generated using 2LPT in ExSHalos; the second row shows the final halo density catalogues using all halos; and the third row shows the final halo density catalogues using all halos with more than $100$ particles, where the abundances of both methods in Fig.~\ref{fig:Halo_Abundance_one} agree with the prediction from Tinker's fitting formula. 

    In the first row of Fig.~\ref{fig:Halo_Maps_one} we see that LPT displacement can reproduce the clumps and filaments seen in the simulation, although they appear fuzzier. 
    This occurs because by using LPT we do not take into account the full non-linear matter evolution, and as a result the particles are less clustered. Therefore, matter halos are larger in LPT displaced catalogues, implying that the usual parameters of halo finders cannot be used in this case.  This adds at least one free parameter to methods that find halos in density maps generated using LPT. Despite having incorrect sizes and the issue with the density of the clumps, the positions of halos are close in both panels, which justifies the use of LPT to displace halos centers.
    
    Therefore, LPT works as a perturbative map between Lagrangian and Eulerian spaces that is fully determined by gravitational collapse. The LPT map recovers well the displacement of density peaks, although it is not sufficient to recover the displacement of the matter around those peaks. 
    
	In the second row of Fig.~\ref{fig:Halo_Maps_one}, we can see the difference between the halo density map of the two methods (simulation and ExSHalos). We see that ExSHalos has many more halos, and describes the matter density map with higher resolution. This could be used to generate galaxy maps using low resolution halo catalogues, saving computational resources.
	
	In the third row of Fig.~\ref{fig:Halo_Maps_one}, we compare halos from the simulation and from ExSHalos, displaying only halos with more than $100$ particles, where the abundances agree with the fitting formula of \cite{Tinker}. We can see that both methods have approximately the same number of halos, but the simulated catalog presents a more clustered map, with clearer filaments and clumps. This makes it explicit the fact that the halo catalogues generated with our method lack power on non-linear scales. Despite this lack of power, the halo density map from ExSHalos can be regarded as a fuzzy version of the fully simulated map, similarly to the maps shown in the first row.
    
    In summary, LPT recovers the cosmic web with less clustered matter and gives the correct positions to the larger halos. As our code only uses LPT to move the halo centers, we do not need any free parameter to define halos in the density field, since this definition is natural in the Gaussian field. Differently from N-body simulations, our method does not display a lack of small mass halos near the resolution limit.


\section*{Acknowledgments}
We thank Francisco Prada for discussions and comments on the draft and Elena Sarpa for help with the LPT implementation. 
RV is supported by FAPESP. 
ML and LRA are partially supported by FAPESP and CNPq.
 
\bibliographystyle{mnras}
\bibliography{main}

\bsp	
\label{lastpage}
\end{document}